\documentclass[preprint]{aastex}

\shorttitle{Non--LTE treatment of molecules}

\newcommand{\dms}{M~dwarfs}
\newcommand{\dls}{L~dwarfs}
\newcommand{\dts}{T~dwarfs}

\newcommand{\phx}{{\scshape phoenix}}
\newcommand{\phoenix}{\phx}
\newcommand{\water}{H$_2$O}

\newcommand{\eqa}{\begin{eqnarray}}
\newcommand{\aqe}{\end{eqnarray}}

\newcommand{\acsul}{{\sum\limits_{l\in L \atop u \in U}}}

\newcommand{\acsuc}{{\sum\limits_{l\in L}}}

\newcommand{\emfactor}{\left(\frac{2hc^2}{\lambda^5}+J_{\lambda}(\lambda)\right)
   {\rm e}^{-hc / \lambda kT} \lambda {\rm d}\lambda}

\newcommand{\absfactor}{J_{\lambda}(\lambda)\lambda {\rm d}\lambda}
\newcommand{\ltefrac}[2]{\frac{n_{#1}^*}{n_{#2}^*}}
\newcommand{\ltefractext}[2]{(n_{#1}^* / n_{#2}^*)}
\newcommand{\fourpi}{\frac{4\pi}{hc}}

\newcommand{\altexp}[1]{{\rm e}^{#1}}

\newcommand{\teff}{\ensuremath{T_{\rm eff}}}
\newcommand{\logg}{\ensuremath{\log (g)}}

\newcommand{\eqref}[1]{(\ref{#1})}
\newcommand{\ang}{\AA}			

\begin{document}

\singlespace

\title{Non--LTE treatment of molecules in the photospheres of cool stars}

\author{Andreas Schweitzer\altaffilmark{1} and Peter H. Hauschildt}
\affil{Dept. of Physics \& Astronomy and Center for 
Simulational Physics, University of Georgia,
    Athens,~GA~30602-2451}
\email{andy@physast.uga.edu}

\and

\author{E. Baron}
\affil{Dept. of Physics \& Astronomy, University of Oklahoma, 
Norman, OK 73091-2061}

\altaffiltext{1}{previous address : Landessternwarte, K{\"o}nigstuhl, 69117~Heidelberg, Germany}

\begin{abstract}
We present a technique to treat systems with very many levels, like molecules, in non--LTE.
This method is based on a superlevel formalism coupled with rate operator splitting.
Superlevels consist of many individual levels that are assumed to be
in LTE relative to each other.
The usage of superlevels reduces the dimensionality of the rate equations
dramatically and, thereby, makes the problem
computationally more easily treatable.
Our superlevel formalism retains maximum accuracy by using
direct opacity sampling (dOS) when calculating the radiative
transitions and the opacities.
We developed this method in order to treat molecules in cool
dwarf model calculations in non--LTE.
Cool dwarfs have low electron densities and a radiation
field that is far from a black body radiation field,
both properties may invalidate the conditions for the common LTE approximation.
Therefore, the most important opacity sources, the molecules,
need to be treated in non--LTE.
As a case study we applied our method to carbon monoxide.
We  find that our method gives accurate results since the conditions
for the superlevel method are very well met for molecules. 
Due to very high collisional cross sections with hydrogen,
and the high densities of H$_2$
the population of CO itself shows no significant deviation from LTE.

\end{abstract}


\section{Introduction}

Frequently, the atmospheres of late type dwarfs are considered to
meet the conditions in which the approximation of
local thermodynamic equilibrium (LTE) can be applied.
In LTE
the population densities of atoms or molecules can be described by the TE
(thermodynamic equilibrium)
Boltzmann--Saha distribution, leading to very simple calculations.
LTE requires that
the collisional rates are large enough
to fully compensate for the deviations of the radiative rates from
their Planckian thermodynamic equilibrium values.
Any spectrum that is not 
a Planckian radiation field
will drive a system
into non--LTE unless the collisional rates are strong enough
to prevent this.
In the extremely cool atmospheres of \dms, \dls\ or \dts , however,
there are two important effects which can lead to deviations
from LTE.
The first is that the extremely low electron densities
and consequently low collisional rates might not be
sufficient to restore LTE.
Collisions with other particles might be much less effective because
of their much smaller relative velocities and their smaller
cross sections.
Investigations for stars like the sun showed that electron
collisions with CO molecules are negligible compared to
all other excitation and de-excitation processes \citep[e.g.][]{thompson73}.
For cooler stars, the electron collisions are even less important
for all molecules, not only CO.
\citet{thompson73} also investigated the effects of collisions with H, H$_2$ and
He on the CO molecule and found them to be important under certain circumstances.
The very detailed study by \citet{ayres89} collected the various existing
collisional cross sections of CO with  H, H$_2$ and He. They found
CO to be in LTE in solar like and cooler stars due to the very high
quasi resonant cross section with H.

Other molecules might be less affected by
atomic collisions.
In general, values for collisional cross sections are only
known very roughly through the formulae of \citet{drawin61}, \citet{vr62}
or  \citet{allen_aq}.
Without good data on collisional excitation cross sections
we cannot be certain that they will dominate the population
distribution in molecules.

In addition, with decreasing effective temperature, the maximum of the energy
distribution stays at roughly 1--1.2 \micron\  \citep{MDpap} since
the strongest opacity sources, TiO in the optical and H$_2$O in the 
infrared, leave only a small window between 1--1.2 \micron\ with relatively
little absorption.  This means that
not only the shape of the spectra are far from spectra of black bodies for
$T=\teff$, but also that
the maximum deviates strongly from the maximum of a black body and the
radiative transitions
``see'' a much hotter local radiative temperature compared to the kinetic
temperature.
This circumstance is particularly important for cool stars which have
broad molecular absorption bands compared to narrow atomic lines for hotter stars.
Of course, for both cool and hot stars, the radiation field produces non-Boltzmann populations
if it is diluted due to optically thin layers, regardless of its shape.

With these different temperatures the radiative rates and the collisional rates
will try to populate the levels
differently.
Non--LTE effects for atomic species such as Ti have already been
investigated by \citet{tinlte,IAU176}. That means that the general conditions for
non--LTE effects (sufficiently low collisional rates and a sufficiently non--Planckian radiation field) are met in
M~dwarf atmospheres.
Since molecules are the dominating opacity sources in cool objects,
deviations from
LTE can have a significant impact on the atmospheric structure and the spectra.

These reasons lead us to develop a method to treat molecules in detailed non--LTE.
We want to emphasize that it is quite well possible that certain molecules
will be in LTE under certain atmospheric conditions (as discussed
above for CO when dominated by collisions with atoms and other molecules).
However, the aim of this paper is to demonstrate how it is possible to
calculate non--LTE effects in molecules in an accurate fashion and to
lay the foundations for calculating molecules that are not in LTE in
cool atmospheres (e.g. molecules that are insensitive to collisions or
molecules that form their spectra in a part of the atmosphere that cannot sustain
the conditions for LTE).

Molecules have many more levels as well as many more lines compared to atoms.
Several thousand levels and several hundreds of thousand lines
are {\em minimal} numbers for very simple molecules. Complicated molecules
(such as \water) will have orders of magnitudes more levels and lines.
The latest line data available for \water\ \citep{ames-water-new} and TiO
\citep{ames-tio} contain 330 and 140 million lines
based on about 50 and 20 million levels respectively.
Since the system of rate equations (equation \eqref{rateeq}, below) is a system
with a rank of at least the number of involved levels, it is obvious that
a straight forward treatment of molecular non--LTE will exceed the limits
of even the largest modern parallel super-computers.

The huge number of lines and levels in molecular systems
is merely a technical difference compared to atomic systems.
There is, however, also an important  physical difference. Atomic processes only
include excitation and ionization processes (and their inverse processes).
In addition, molecular processes include dissociation and formation of
the molecule through different possible channels.
That means that a realistic model has to include not only
the molecule itself but also take into account
the constituting atoms and the equation of state has to be
aware of a possible source or sink of atoms.
The logical extension is ``Non Local Chemical Equilibrium (NLCE)''.
As a first step, we address in this paper only internal non--LTE
and do not account for dissociation.
Neglecting NLCE is a good approximation
for very stable molecules which are not easily dissociated
since the kinetic and radiative energies in M~dwarf or cooler
atmospheres are too small to do so.

So far only very little work has been done on general molecular non--LTE
mainly because of the difficulties mentioned above.
Usually, only very few levels close to the ground state are considered.
CO, however, has experienced a number of investigations.
The limiting factor for CO is the uncertainty in the quality of the very
high collisional rates.
A recent and important example is the work by Ayres and Wiedemann \citep{wiedemann94,ayres89}
who calculated CO in non--LTE in order to try to explain the presence
of strong CO bands in the solar infrared spectrum,
which cannot be explained by the high kinetic energies of
the gas in the solar atmosphere.
They only used 10 vibrational levels (which produce the optically thick
lines) and the rotational levels were
simplified in a fashion similar to what will be presented here, i.e.
they assumed some LTE behavior between certain levels.

Another example is the work by Kutepov et al. \citep{kutepov97,kutepov91},
who calculate the solar radiation in the earth's atmosphere through molecular
bands including CO. They consider only
three vibrational levels and started with so called vibrational LTE (which is also
conceptually very similar to the work presented here) but relaxed this assumption
in their latest work.

In the next section we provide the necessary background for this work.
We review the atmosphere code used and the physics involved.
In section \eqref{molnlte_super} we present the superlevel
method on which this paper is based on.
In section \eqref{co_nlte} we apply our method to carbon monoxide.
We use CO as an example and demonstrate the numerical
implementation.
We conclude our paper in section \eqref{concs}.

\section{Background}

\label{back_phoenix}
This work is based on the general stellar atmosphere code \phx\
described in \citet{fe2effects,novaphys97,phxparallel,phxpar98,NGhot}.

\label{back_nlte}

The solution of the wavelength dependent radiative transfer equation is performed by
operator splitting
as described in 
\citet{s3pap} where
the  mean intensity $J$ is iteratively calculated via
\begin{equation}
\label{lambda_def}
J_{\rm new}=\Lambda^* S_{\rm new}+(\Lambda-\Lambda^*)S_{\rm old}.
\end{equation}
where $S$ is the source function, $\Lambda$ the lambda
operator expressing the formal solution and $\Lambda^*$ a suitably
chosen approximate lambda operator.

One major
feature is that \phx\ calculates each
wavelength point separately, i.e. \phx\ steps through the
wavelength grid and solves the radiative transfer equation
at each wavelength point by only using 
data important 
for that wavelength point.
This dramatically reduces the memory requirements for the code.
Only the quantities that need to be integrated over
wavelength are kept in memory.

Another important
feature to be mentioned here is the use of line lists
which contain several hundred million spectral lines.
\phx\ has been especially optimized to handle this task
by using  dynamical opacity sampling (dOS)
\citep[see also][]{phd,cygpap}.
dOS dynamically discards the negligible lines for
a particular atmosphere configuration and does not 
require precomputed tables.
With this feature \phx\ can calculate spectra
with high resolution and a large number of spectral
lines efficiently.
The superlevel method presented below
will take great advantage of 
dOS as it will improve the accuracy and reduce the necessary
approximations.

In order to calculate the population densities of individual levels in molecules
(or atoms) without the assumption of thermodynamic equilibrium,
we need to solve the rate equations which balance
all population and de-population processes for every level.

The radiative absorption rates for individual levels
due to absorption of photons are described by
the Einstein $B_{lu}$ coefficient\footnote{
The indices used throughout the paper have the following meanings~:
$l$ always stands for
a lower level, $u$ always stands for an upper level and $i$ or $j$
are used where a distinction between upper and lower level is
not possible or not necessary.
}
and the profile averaged mean intensity $\bar{J_{ul}}$ 
or by the
cross section $\alpha_{lu}^{\rm abs}$ and the mean intensity $J$~:
\begin{equation}
\label{abscoeff}
R_{lu} = B_{lu}\bar{J_{ul}} = \fourpi \int_0^{\infty} \alpha_{lu}^{\rm abs} \absfactor
\end{equation}
The emission rates are
the sum of induced emission and spontaneous emission. However, to 
simplify the rate equations we shall follow \citet{mihalas78} and use
a radiative emission rate coefficient $R_{ul}$ between individual levels
\begin{equation}
\label{emcoeff}
R_{ul}=\fourpi \int_0^{\infty} \alpha_{ul}^{\rm em} \emfactor
\end{equation}
such that we can write the emission rate as
$
n_u ( n_l^* / n_u^* ) R_{ul}.
$

With these radiative rate coefficients and with the collisional rate
coefficients $C_{ul}= \ltefractext{l}{u}C_{lu}$ the rate equation for one
particular level $i$ can be written as
\begin{equation}
\label{rateeq}
         \sum_{l<i} n_l                        (R_{li} + C_{li})
   +     \sum_{u>i} n_u       \ltefrac{i}{u}  (R_{ui} + C_{iu})
 =       n_i \Big( \sum_{u>i}                  (R_{iu} + C_{iu})
   +     \sum_{l<i}           \ltefrac{l}{i}  (R_{il} + C_{li}) \Big).
\end{equation}
The system of all rate equations
is closed by the conservation equation
for the particles and the charge conservation equation. 
For a multi level system, equation \eqref{rateeq} is a non-linear system
of equations with 
respect to
the population densities since the radiative rate coefficients 
themselves depend on the population densities via the mean intensities.
It is also non-linear with respect to the electron density via
the collisional rate coefficients and the charge conservation
equation.
In the case of molecules that can dissociate and form through various
channels the situation will get more complicated since
system \eqref{rateeq} will include not only the molecule
in question but also all atoms and all possible molecules which have to be
calculated simultaneously.
Furthermore, the condition of particle conservation has to be exchanged for
the condition of nuclei conservation.

We use the same formalism as \citet{casspap} and solve equation \eqref{rateeq}
with rate operators and operator
splitting techniques.
Then the rates can be written by means of a rate operator $[R_{ij}]$
and a population density operator $[n_i]$~:
\begin{equation}
R_{ij}=[R_{ij}][n_i]
\end{equation}
The radiative rates are functions of the mean intensity $J$ which
is usually expressed by the lambda operator and the source
function $J=\Lambda S$.
Therefore, the rate operator is also a function of the
lambda operator.
However, since the source function is non-linear with respect to
the population densities, the lambda operator is substituted
by the $\Psi$ operator introduced by \citet{rh91}
\begin{equation}
\Lambda(\lambda)=\Psi(\lambda)\left[\frac{1}{(\kappa_{\lambda} + \sigma_{\lambda})}\right].
\end{equation}
This will eliminate the full source function and introduce
the emissivity $\eta_{lu}(\lambda)$ instead, which can be expressed by the operator $[E(\lambda)]$ by
\begin{equation}
[E(\lambda)][n]=\sum_{l<u}\eta_{lu}(\lambda)+\tilde{\eta}(\lambda)
\end{equation}
where $\tilde{\eta}(\lambda)$ is the background emissivity.

With these definitions, the absorption rates in eq. \eqref{abscoeff} can
be expressed as
\begin{equation}
[R_{lu}][n] = \fourpi \left[ \int_0^{\infty} \alpha_{lu}^{\rm abs} \Psi(\lambda) E(\lambda)
\lambda {\rm d}\lambda \right] [n]
\end{equation}
and the emission rate from eq. \eqref{emcoeff} becomes
\begin{equation}
[R_{ul}][n]=\fourpi \int_0^{\infty} \alpha_{ul}^{\rm em} 
\left(\frac{2hc^2}{\lambda^5}+\Psi(\lambda) [ E(\lambda)] [n] \right)
{\rm e}^{-hc / \lambda kT} \lambda {\rm d}\lambda
\end{equation}

In order to solve the system of rate equations with an 
operator splitting iteration scheme,
the rate operators are split in analogy to the approximate lambda
operator \eqref{lambda_def}~:
\begin{equation}
R_{ij}=[R_{ij}^*][n_{\rm new}]+([R_{ij}]-[R_{ij}^*])[n_{\rm old}]
=[R_{ij}^*][n_{\rm new}]+[\Delta R_{ij}][n_{\rm old}].
\end{equation}
The rate equations \eqref{rateeq} can then be written in the form
\begin{eqnarray}
\label{rateeq_op}
      &   &   \sum_{l<i} n_{l,{\rm new}}                [R_{li}^*][n_{\rm new}]
            + \sum_{u>i} n_{u,{\rm new}} \ltefrac{i}{u} [R_{ui}^*][n_{\rm new}] \nonumber \\
      & + &   \sum_{l<i} n_{l,{\rm new}}                ([\Delta R_{li}][n_{\rm old}] + C_{li})
            + \sum_{u>i} n_{u,{\rm new}} \ltefrac{i}{u} ([\Delta R_{ui}][n_{\rm old}] + C_{iu}) \nonumber \\
      & = &   n_{i,{\rm new}} \Big( \sum_{u>i}          [R_{iu}^*][n_{\rm new}] 
            + \sum_{l<i}                 \ltefrac{l}{i} [R_{il}^*][n_{\rm new}]  \Big) \nonumber \\
      & + &   n_{i,{\rm new}} \Big( \sum_{u>i}          ([\Delta R_{iu}][n_{\rm old}] + C_{iu})
            + \sum_{l<i}                 \ltefrac{l}{i} ([\Delta R_{il}][n_{\rm old}] + C_{li}) \Big).
\end{eqnarray}
This equation can be solved iteratively and represents an 
operator splitting
iteration analogous to solving the radiative transfer
equation. 

In \phx, equation \eqref{rateeq_op} is solved after having solved
the radiative transfer equation for all wavelength points.
During the wavelength loop all the rates and operators are integrated
using the newly calculated mean intensity.
Convergence of the occupation numbers is obtained by iterating this calculation with
updated occupation numbers (and a possible temperature correction
applied) until overall convergence.

\section{The superlevel formalism}

\label{molnlte_super}

\subsection{General concept}
For complex molecules, the dimension of eq.~\eqref{rateeq_op} 
becomes very large (see sec.~1).
However, the memory requirements for a computer program that has to solve eq.~\eqref{rateeq_op}
with a direct approach will scale with the number of levels squared.
Therefore, depending on the molecular system and the available computer resources
the problem of solving the rate equations becomes untreatable and needs to be simplified.
One possibility to simplify non--LTE problems is to treat
the system ``partially'' in LTE. This was done, e.g., by Kutepov et al. and
Ayres and
Wiedemann as mentioned above. They assumed LTE population within one vibrational
state.

A much more general approach was presented by \citet{anderson89} who faced the
problem of complex atoms and ions with a lot of levels (e.g.
\ion{Fe}{2}, \ion{Ni}{2} etc.).
He also grouped several levels into one common level and assumed LTE within
such a group. However, the exact grouping was relaxed to ``some similarity'' of 
all levels within one group. Also, he derived methods how to calculate
the opacity by such levels by means of opacity distribution functions.
This idea was also the basis of the work by \citet{dreizler93} and by \citet{hubeny95}.

With the rapid development of large computers it is no longer
necessary to 
use superlevels for atomic systems.
\citet{novaphys97} and \citet{fe2effects}, e.g., calculate
large atomic systems directly in non--LTE.
For molecules, however, the number of levels is orders
of magnitudes larger and approximations are still
necessary. Furthermore, these approximations
are much better for molecules than they are for atoms.
The typical energy differences between individual molecular
levels are much smaller than the energy differences between
atomic levels. This provides very good internal coupling
and thermalization within one superlevel.

We will only adopt the basic idea of superlevels, that is, we group many individual
levels into one superlevel and assume LTE populations within one superlevel.
The populations of the superlevels will then be calculated via
the rate equations.
However, the absorption coefficients and the transition rates between the superlevels will not be approximated.
We will account for all individual lines by dynamically sampling the opacity and
the transition rates.
This distinguishes our approach from the original approaches mentioned above.
There, it had been necessary to develop opacity distribution functions for the
transitions between the superlevels. This is not needed here.
As explained above, \phx\ uses dOS and, therefore, a dynamical opacity sampling
of transitions between superlevels is already done by design.
The only purpose of the superlevel formalism is to keep the system of rate
equations small.

A superlevel\footnote{
In the following, the same nomenclature conventions apply as in section \eqref{back_nlte}
with the addition, that capital letters like $I$, $L$ or $U$ are used for superlevels, 
and that lower case
letters like $i$, $u$ or $l$ are used for actual levels.
}
$I$ is constructed out of a number of actual levels $i$ and
its population number density is defined by
\begin{equation}
 n_I = \sum_{i \in I} n_i. 
\end{equation}
For the population densities of the actual levels within one
superlevel LTE is assumed and the Boltzmann equation can be used
for a level $i\in I$~:
\begin{equation}
\label{lteratio}
 n_i = n_I \frac{g_i}{Z_I} {\rm e}^{-E_i/kT} 
\end{equation}
where 
\begin{equation}
 Z_I = \sum_{i \in I} g_i {\rm e}^{-E_i/kT} 
\end{equation}
is the {\em finite} partition sum over the superlevel considered.

Eq. \eqref{lteratio} can be used for both the LTE and the
non--LTE quantities.
With this we define the departure coefficients as in \citet{menzel37}
and \citet{mihalas78} as
\begin{equation}
b_i = \frac{n_i}{n_i^*} = \frac{n_I}{n_I^*} = b_I,
\end{equation}
where the $n_i^*$ are the occupation numbers calculated via the
Saha Boltzmann equation and the actual non--LTE population density
of the continuum.
Also, in LTE equation \eqref{lteratio} holds true for {\em any} level
$i\notin I$. Summing equation \eqref{lteratio} over all levels $i\in J$ 
we get
the ratio $n_I^*/n_J^*$ of two superlevels 
to~:
\begin{equation}
\label{sulteratio}
\frac{n_I^*}{n_J^*}=\frac{Z_I}{Z_J}
\end{equation}
which is the analogous relation to the Boltzmann relation
$n_i^*/n_j^*=g_i/g_j\exp (-E_{ij}/kT)$ for 
normal levels and is in fact only a more general form
of equation \eqref{lteratio}.

\subsection{The partition functions}

Equation (\ref{sulteratio}) can be written as
\begin{equation}
\frac{n_J}{n_I^*}= \frac{n_J}{n_J^*} \frac{Z_J}{Z_I}
\end{equation}
Summing over all superlevels and adding all remaining levels
not included in any superlevel yields
\begin{equation}
n_I^* = n_{\rm tot} \frac{Z_I}{\sum_I b_I Z_I + \sum_{i \not\in I}
b_i g_i \exp(-E_i / kT)}
\end{equation}
This leads to the definition of the non--LTE partition function which can
be written for superlevels
\begin{equation}
\label{nlte_q_def}
Q_{\rm non-LTE}^S =\sum_I b_I Z_I + Q_{\rm corr}
\end{equation}
where $Q_{\rm corr}$ is the sum over all levels not included in
any superlevel.
By using the definition of $Z_I$ and $b_i=b_I$ this can be expanded
to the usual non--LTE partition function
\begin{equation}
\label{nlte_part_func}
Q_{\rm non-LTE}^S = \sum_I {\textstyle \sum \limits_{i \in I}}
b_i g_i {\rm e}^{-E_i/kT} + \sum_{i \not\in I}
b_i g_i {\rm e}^{-E_i/kT} = \sum_{i=0}^{\infty}
b_i g_i {\rm e}^{-E_i/kT}
\end{equation}
With these relations it is possible to obtain LTE and non--LTE
occupation numbers from the departure coefficients.
If we had used  the  definition of departure coefficients introduced by
\citet{zwaan72}
the LTE occupation numbers
could be obtained simply by solving the Saha--Boltzmann
equations for the complete molecule.
However, we use the definition by \citet{menzel37}
and, therefore, it is necessary to use the partition
function \eqref{nlte_q_def}.

Generally, the partition function is an infinite sum as seen in
equation \eqref{nlte_part_func} which mathematically does not converge.
Physically, this sum is ``cut off'' at some
level which can never be populated due to the presence of
other neighboring particles \citep[see e.g.][]{mihalas78}.
If the number of levels for which data are available is large enough,
it is sufficient to set $Q_{\rm corr}=0$ and circumvent
the problem of deciding how to cut off the infinite LTE partition function.

\subsection{The radiative  rates within the superlevel formalism}

\label{superradrates}

The emission rate between two superlevels is the sum of all emissions $ n_u(n_l^*/n_u^*)R_{ul}$
between all relevant actual levels.
Inserting $(n_U n_U^* n_L^*)/(n_U n_U^* n_L^*)$ and expressing
$R_{ul}$ by the cross section $\alpha_{ul}^{\rm em}$ of individual 
lines (equation \eqref{abscoeff}) yields :
\[
 \sum\limits_{u\in U\atop l \in L} n_u \frac{n_l^*}{n_u^*} R_{ul}
            =  n_U \frac{n_L^*}{n_U^*} \fourpi
               \int_0^{\infty} \acsul  \frac{n_u}{n_U} \frac{n_U^*}{n_L^*} \frac{n_l^*}{n_u^*}
               \alpha_{ul}^{\rm em} \emfactor
\]
and when using equations \eqref{lteratio} and \eqref{sulteratio}
\begin{eqnarray}
 \sum\limits_{u\in U\atop l \in L} n_u \frac{n_l^*}{n_u^*} R_{ul}
          & = & n_U \frac{n_L^*}{n_U^*} \fourpi
               \int_0^{\infty} \acsul  \frac{g_l}{Z_L} {\rm e}^{-E_l/kT}
               \alpha_{ul}^{\rm em} \emfactor \nonumber \\
 \label{suemcoeff}
          & = & n_U \frac{n_L^*}{n_U^*} \, R_{UL} 
\end{eqnarray}
 where
 \begin{eqnarray}
 \label{suemrate}
 R_{UL}               & \equiv &  \fourpi \int_0^{\infty} \alpha_{UL}^{\rm em}\emfactor \\
 \label{alphaemdef}
 \alpha_{UL}^{\rm em} & \equiv &  \acsul \alpha_{ul}^{\rm em} \frac{g_l}{Z_L} {\rm e}^{-E_l/kT}
 \end{eqnarray}
Now, the rate coefficient \eqref{suemrate} has the same form
and properties (see equation \eqref{suemcoeff}) as
the rate coefficient \eqref{emcoeff} for  ``normal'' levels.

Similarly, the absorption rate between two superlevels is the sum over all relevant
absorptions $n_l R_{lu}$
and becomes
 \begin{equation}
   \label{suabsrate}
    n_L R_{LU} = n_L  \left \{ \fourpi \int_0^{\infty}\alpha_{LU}^{\rm abs} \absfactor \right \}
 \end{equation}
 where 
  \begin{equation}
    \label{alphaabsdef}
  \alpha_{LU}^{\rm abs} \equiv    \acsul \alpha_{lu}^{\rm abs} \frac{g_l}{Z_L} {\rm e}^{-E_l/kT}
  \end{equation}
 Again, the definition of $R_{LU}$ has the same form as the
 equivalent expressions \eqref{abscoeff} for actual levels.

Note that equations \eqref{alphaemdef} and \eqref{alphaabsdef}
only differ in the choice of $\alpha^{\rm em}$ or
$\alpha^{\rm abs}$ and that in the case of complete
redistribution $\alpha_{UL}^{\rm em}$ and $\alpha_{LU}^{\rm abs}$
are identical also for superlevels. 
That means that a computer program needs to keep track of only {\em one}
quantity which is less
memory or time consuming 
than keeping track of the sum over the absorption cross sections {\em and}
the sum over the emission cross sections (note
that the cross sections need to be known for
every depth point of an atmosphere and for every transition).

With the two radiative rates \eqref{suemrate} and \eqref{suabsrate}
it is possible
to calculate all the radiative rates needed to solve the rate
equation \eqref{rateeq}.
The rates have to be obtained via wavelength integration during the
wavelength loop (as described in section \eqref{back_phoenix})
and depend on all actual transitions between
two {\em different} superlevels.
Since all {\em actual} transitions have to be calculated
this does not seem to simplify the original problem of
dealing with millions of lines.
However, \phx\ is already designed to calculate as many transitions
as desired very efficiently, as explained in section
\eqref{back_phoenix} by means of dOS. Therefore, it is
straight forward to calculate the radiative rates ``as exactly
as desired''.
The computer program simply has to assign each actual transition
to the super transition it belongs to.
Note also, that the purpose of the superlevels in this approach is to
reduce the number of levels in the rate equations 
not the number of lines calculated.
This keeps the accuracy of the rate coefficients and of the absorption coefficients
as high as possible.
Only the population densities will be approximated.
It also keeps the resulting spectrum as accurate as possible.

The rate equations are solved by operator splitting techniques as described
in section \eqref{back_nlte}.
Note that the operator technique only relies on the $\Psi$ operator and
the $E$ operator. The $\Psi$ operator is derived from the $\Lambda$ operator
and the $E$ operator is derived from the emissivity.
The $\Lambda$ operator obviously does not depend on the superlevel
formalism. The $E$ operator depends on the opacity, but as already 
explained, the treatment of the opacity is not changed
by the superlevel formalism.

\subsection{The absorption and emission coefficients}

As mentioned above,
the only purpose of the superlevels is to keep the system
of the rate equations small. There is no need to keep the number
of transitions small.
Therefore, every absorption or emission coefficient
is calculated between actual levels.
As explained above, \phx\ uses lists of spectral lines as
input data. The $gf$ values (and other data) for the systems treated with
superlevels can be (and are) included in the same input data.
Therefore, the cross sections
$\alpha_{lu}$ are available for every individual transition at
every wavelength point selected by dOS as explained above.
Care has to be taken when entering 
the number density into
the absorption coefficient:
equation \eqref{lteratio} has to be applied to obtain
the number density of a particular actual level
out of the number density of the superlevels.

That means that,
as explained for the radiative rates, our computer code calculates
``exact'' cross sections $\alpha_{LU}$ for superlevel transitions
with complicated
``superline profiles'' and does not require 
opacity distribution functions for the supertransitions.
The advantage 
regarding the absorption and emission coefficients and
the spectrum synthesis
is that we can calculate a very accurate
spectrum at any desired resolution within our superlevel
formalism.

\subsection{Collisional and continuum rates within the superlevel formalism}

The collisional excitation rate from one superlevel to another superlevel is the sum of
all excitation rates between the relevant actual levels
and becomes
 \begin{equation}
 \label{sucollabs}
 n_L C_{LU} = n_L \left\{ \acsul \frac{g_l}{Z_L}\altexp{-E_l/kT} C_{lu} \right\}.
 \end{equation}

Similarly, the rates for collisional de-excitation between two superlevels
are :
\begin{equation}
\label{sucollem}
n_U C_{UL} = n_U \left \{ \acsul \frac{g_l}{Z_U}\altexp{-E_l/kT} C_{lu}  \right \}
\end{equation}

With the definitions \eqref{sucollabs} and \eqref{sucollem} and the relation
\eqref{sulteratio}, it follows immediately that
\begin{equation}
\label{sucollratio}
\frac{C_{LU}}{C_{UL}} = \frac{Z_U}{Z_L} = \frac{n_U^*}{n_L^*}
\end{equation}
which is the equivalent to 
the Boltzmann relation for the collisional rate coefficients for actual levels.

With equations \eqref{sucollabs}, \eqref{sucollem} and \eqref{sucollratio} it is now
possible to calculate the collisional rates $n_I C_{IJ}$ needed
for the rate equation \eqref{rateeq} from known
collisional cross sections and coefficients $C_{lu}$.

Ionization and recombination processes can be included very
similarly.
For photoionization
\begin{equation}
             n_L R_{Lc} =  n_L \left\{ \fourpi \int_0^{\infty}\alpha_{Lc}^{\rm ion} \absfactor \right\} \\
\end{equation}
where
\begin{equation}
  \alpha_{Lc}^{\rm ion} \equiv    \acsuc \alpha_{lc}^{\rm ion} \frac{g_l}{Z_L} {\rm e}^{-E_l/kT}
\end{equation}
and for radiative recombination

\begin{equation}
           n_c \frac{n_L^*}{n_c^*} \, R_{cL}  =  n_c \frac{n_L^*}{n_c^*} \, \left \{ \fourpi \int_0^{\infty} \alpha_{cL}^{\rm recomb}\emfactor \right \} \\
\end{equation}
where
\begin{equation}
\alpha_{cL}^{\rm recomb} \equiv   \acsuc \alpha_{cl}^{\rm recomb} \frac{g_l}{Z_L} {\rm e}^{-E_l/kT}
\end{equation}
For the collisional ionization
\begin{equation}
            n_L C_{Lc} = n_L \left \{ \acsuc \frac{g_l}{Z_L}\altexp{-E_l/kT} C_{lc} \right \}
\end{equation}
and for the collisional recombination
\begin{equation}
            n_c C_{cL} = n_c \left \{ \frac{n_L^*}{n_c^*} \acsuc \frac{g_l}{Z_c}\altexp{-E_l/kT} C_{lc} \right \}
\end{equation}

However, usually the most important continuum processes
of molecules are dissociation and molecular recombination.
Including these processes requires accounting for all particles that
are involved in a molecular production and destruction chain,
i.e. all constituent atoms and all possible
molecules that can be constructed out of these atoms via all
possible chemical pathways.
Such a NLCE (non local chemical equilibrium) treatment
is beyond the scope of this work since it
requires a formalism beyond the superlevel formalism.
Important molecules with low dissociation energies like TiO
could be affected by NLCE.
However, it can be neglected for molecules
that have dissociation energies
less than those that occur
in M~dwarf atmospheres, e.g. for CO.

\section{Implementation of the superlevel formalism for CO}
\label{co_nlte}

We test our method on carbon monoxide (CO).
This diatomic molecule has data available that
are detailed enough and
of good quality \citep[see e.g.][]{diplom}. Also, it does not have
very many lines and levels; and the lines have very distinct bands
in the infrared (most prominently at 2.3~\micron\ and at 4.3--6~\micron).
CO not only has reliable data, but
is also important in other astronomical fields like
the earth's atmosphere and the solar atmosphere.

Other molecules like TiO or H$_2$O have either data which are not 
detailed enough, i.e. the level data cannot be extracted,
or the data are most likely incomplete and not very
accurate \citep{h2olet,ara97}.
Furthermore, CO is a relatively small system. With only a 
few thousand levels (about a factor of 1000 less than TiO
or \water) it is only slightly larger than the largest
atomic system currently calculated in non--LTE.
Therefore, we will compare the superlevel calculations with a direct
non--LTE calculation.
Thus, we can quantify the accuracy of the superlevel
treatment by direct comparison to an ``exact'' calculation.

\subsection{Available Data}

\subsubsection{Molecular line and level data}
The necessary level data, i.e. the excitation energies,
and the quantum numbers as well as the necessary line data, i.e.
the transition wavelengths, the transition strengths and
the involved levels, were extracted from
the line list by Goorvitch et al. \citep{goorCO,goorCOa,goorCOb}.
The line list is detailed enough to uniquely assign 
the upper level to each transition.

For simplicity, only $^{12}{\rm C}^{16}{\rm O}$ was considered.
It contains 3623 levels and 19203 transitions for the electronic
ground state. The first excited electronic state has a very high
energy and all transitions to this state are in the UV. Therefore,
these transitions are not included in the available line list as
these states and transitions are unimportant in \dms.

\subsubsection{Collisional cross sections}

For collisions with electrons the formula given in \citet{allen_aq}
has been used for simplicity.
This is a very rough approximation but electron densities
are very low in \dms\ and as already noted by
\citet{thompson73} or \citet{ayres89}, e.g., electron
collisions  with CO are negligible even for higher 
electron temperatures than in \dms.

For collisions with atomic and molecular hydrogen and with helium we
used the cross sections collected by \citet{ayres89} and expressed in
terms of
\begin{equation}
\Omega_{ul}=4.2 \times 10^{-19}\frac{\exp(B_x-0.069A_x\beta^{1/3})}{\beta(1-e^{-\beta})}
\end{equation}
where $\beta=E_{ul}/kT$ and $\Omega_{ul} = R_{ul}/n_x$.
The subscript $x$ stands for the collisional partner.
$A_x$ and $B_x$ are also taken from \citet{ayres89}.
Explicitly,
$A_x$ is 3, 64 and 87 for H, H$_2$ and He, respectively, and
$B_x$ is 18.1, 19.1 and 19.1 for H, H$_2$ and He, respectively.
These are values originating from the results of \citet{glass82} for H and H$_2$ and
the results of \citet{milikan64} for He.

\subsubsection{Data for continuum processes}
CO has a dissociation energy of 11.09~eV  and an
ionization energy of 14.01~eV.
These energies
are much too high to be achieved in M~dwarf atmospheres
(10 eV correspond to 77\,200~K or 1240~\ang).
Also, no data for detailed cross sections for ionization
or dissociation are available.
Therefore, the molecule is treated in ``quasi--internal non--LTE'', i.e.,
our calculations can only determine the population densities of
the energy levels of CO, not the number density of CO.
That means we neglect NLCE and the dissociation and ionization transitions are accounted for, but only
with negligible transition rates and dissociation processes
are not accounted for in the atomic equation of state.
This will be improved
in future work.

\subsection{The division into superlevels}
\begin{figure}
\plotone{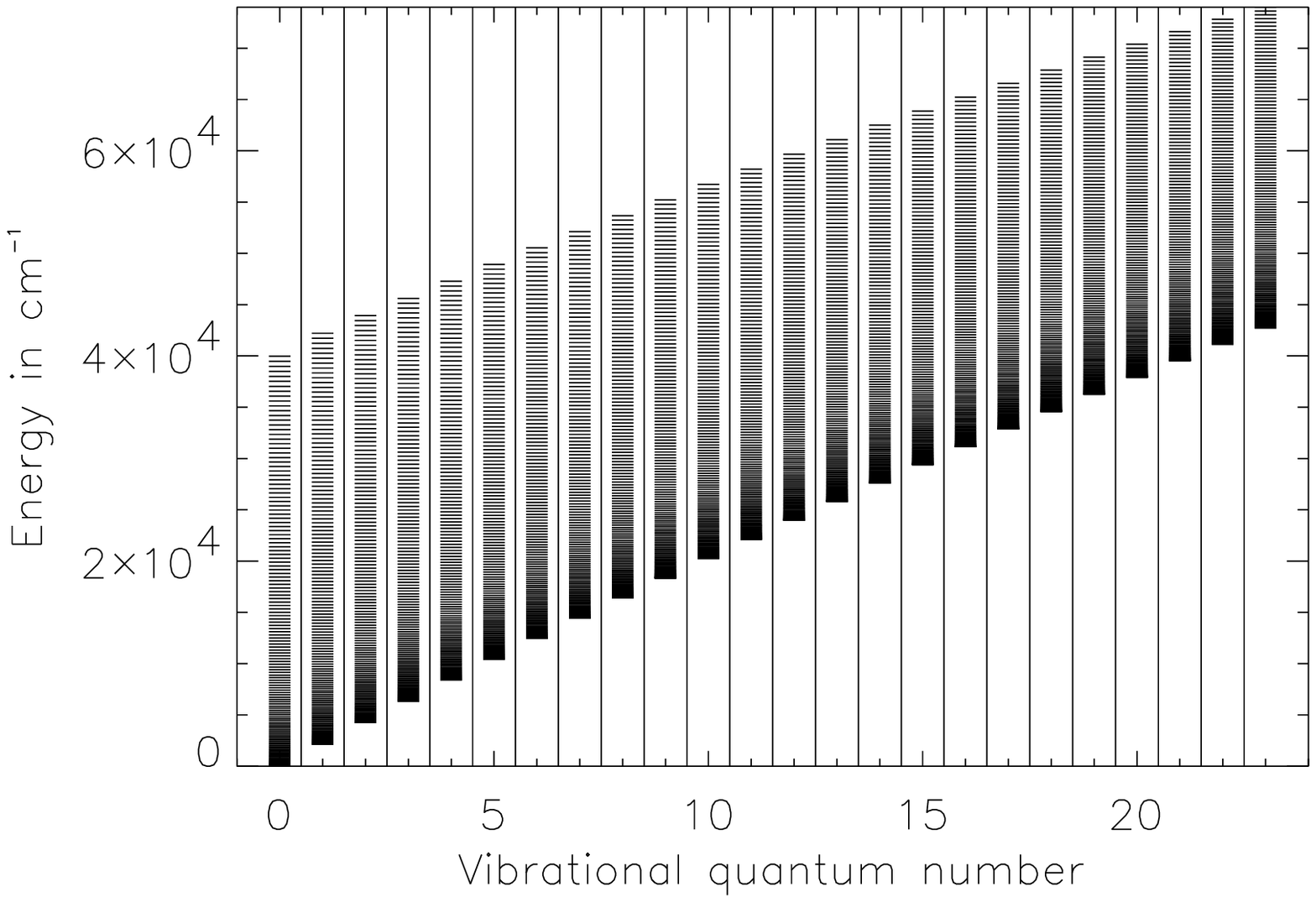}
\caption{\label{co_quant_grot}
Energy level diagram for the electronic ground state of CO.
The bound-bound transitions have been omitted for clarity.
The vertical lines mark the superlevel boundaries
for Model A (see text for details).}
\end{figure}

\begin{figure}
\plotone{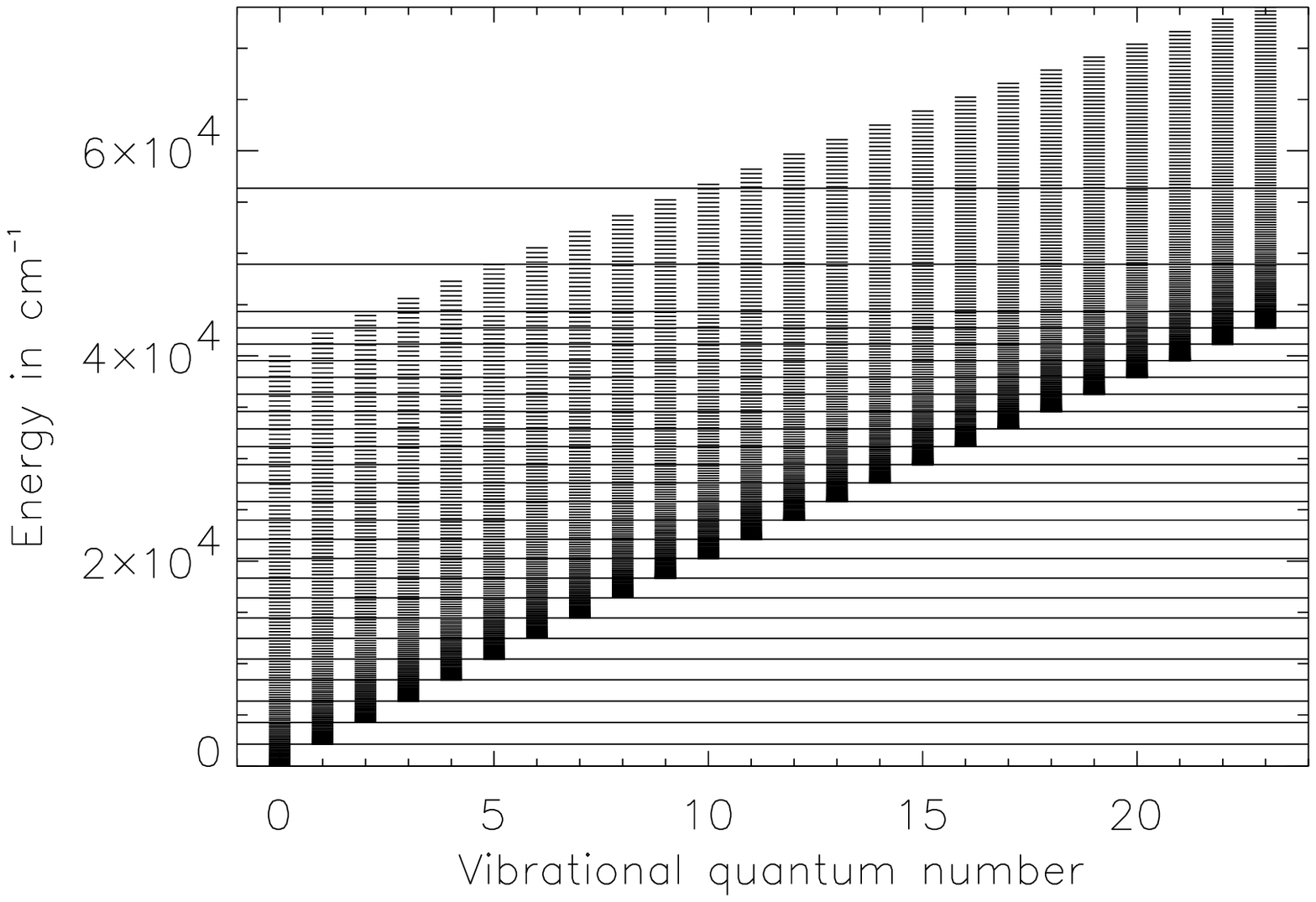}
\caption{\label{co_head_grot}
Energy level diagram for the electronic ground state of CO.
The bound-bound transitions have been omitted for clarity.
The horizontal lines mark the superlevel boundaries
for Model B (see text for details).}
\end{figure}

Energy level diagrams of Carbon monoxide are shown in 
Figs. \eqref{co_quant_grot} and \eqref{co_head_grot}.
The energy levels have been grouped
into vibrational quantum numbers and the transitions have been
omitted for clarity.
When dividing the levels into superlevels
we have to assure that the
coupling within one superlevel is very strong.
Then we can be certain that the requirements for the superlevel approximation are met
and that we can assume that we have a thermal population within one superlevel.
The following possible divisions into superlevels have been considered :
\subsubsection{Model A}
An obvious splitting into superlevels is the
grouping by constant vibrational quantum number,
assuming strong coupling between the rotational
states for one vibrational quantum number.
This assumption is very good since the purely
rotational transitions have very long lifetimes.
The Einstein A values for purely rotational
transitions are of the order year$^{-1}$.
Therefore, collisions will occur much more often
and rotational thermalization can be assumed.

We will call this model ``Model A''.
It corresponds to vibrational LTE as used by
\citet{kutepov91}.
In Fig. \eqref{co_quant_grot},
the vertical lines mark the boundaries for Model~A.

However, this grouping results in a large overlap in energy
between all superlevels and a wide range of energies
affects one superlevel. This may make the thermalization assumption invalid.
\subsubsection{Model B}

The second superlevel model selects
superlevels only by their energy.
The energy boundaries are defined by the rotational
ground states for each vibrational state.
The original authors of the superlevel idea
\citep{anderson89,dreizler93,hubeny95}
also used the energy as a selection criteria for their
atomic systems.
At face value, the pure energy criteria does not seem to meet the
requirements for the superlevel treatment to be valid.
There is, a priori, no strong coupling for levels with
similar energies. For atoms, this criterion
has been a matter of pure feasibility.
In our case, however, each superlevel defined by the energies
of the actual levels will still be dominated by the levels
with one (the highest) vibrational quantum number.
As explained above, it is a good approximation to assume
strong coupling for constant vibrational quantum number.
If all the other levels included in such a superlevel that 
do not have the same vibrational quantum number
are not important for balancing the rate equations,
they do not have to couple strongly to the dominating
levels. 

This model we will call ``Model B''.
Such a superlevel definition can be seen in Fig. \eqref{co_head_grot}
indicated by the horizontal lines which are the 
superlevel energy boundaries.

The vibrational level with the highest
quantum number and the other levels with corresponding energies
would produce a huge and wide spread
superlevel. This superlevel was divided in such a way 
that the resulting superlevels contain roughly the same
number of levels per superlevel as does the average superlevel.

\subsubsection{Model C}

The third superlevel model combines both criteria from
Model~A and Model~B.
That means we select all levels by their vibrational
quantum number and by their energy.
That way we have the advantages of both methods.
The internal coupling can be assured by purely rotational
transitions which only can occur via collisions (as explained
for Model~A) and, in addition, there is no energetical
overlap between the superlevels.
Furthermore, this model represents a step between Models~A or B and
a direct solution without superlevels.
This model has only a factor of ten less superlevels than
actual levels.
Our model contains 350 superlevels for
this method compared to 24 and 27 for Model~A and Model~B,
respectively.

We call this model ``Model C''. It can be visualized by
plotting Fig.~\eqref{co_head_grot} over Fig.~\eqref{co_quant_grot}.
The resulting rectangles represent the superlevel boundaries.

\subsubsection{Model Z}

The last model we consider is direct non--LTE using all available
levels and lines.
The 3623 levels and 19203 lines of the CO model 
are within 
the limits of a modern workstation.
This model we call ``Model Z''.
Note, that the superlevel formalism reduces to ``regular'' non--LTE
when the sums are carried out over only one term.
Therefore, regular non--LTE can easily be treated within
our superlevel formalism.
We will use the results from this model to evaluate our
superlevel formalism.

\subsection{Numerical implementation}
\label{numerical}

The non--LTE implementation for molecules is implemented in a manner very similarly to
the already existing non--LTE implementation for atomic species 
as explained in section \eqref{back_nlte}.
The necessary level data for all actual individual levels
are obtained from the input line list.
The absorption coefficients
for individual CO lines are calculated with the non--LTE occupation
numbers via equation \eqref{lteratio}.
For every wavelength point the cross sections between
superlevels  are obtained as explained
in section \eqref{superradrates} and are integrated over
the spectrum.

In order to select a certain superlevel model
we need only change the input files. The computer code
is independent of assumptions about the model molecule.
This makes it very easy to change between the various
models.

The rate equations are solved as described in
section \eqref{back_nlte} using the operator
splitting technique. Convergence of the departure
coefficients $b_i$ is obtained
simultaneously with the iteration of the temperature structure.

\subsubsection{The accuracy of the rates}

The convergence properties of the rate operator method
depends strongly on the accuracy of the rates.
The accuracy of the rates is determined by the accuracy of
the integration of
the radiative rates.
This has been achieved
by a carefully selected wavelength grid and by correcting
the first order error via normalization of the
line profiles used for the radiative rates.

The wavelength grid on which the spectrum is calculated has to
include all transitions between all superlevels. For the accuracy
of the rates it does not
have to include the transitions that occur within one superlevel.
The wavelength grid has been optimized to sample the molecular
line profiles with as few wavelength points as
necessary.
Otherwise, the number of necessary
wavelength points would be at least a multiple of the number of
individual lines.
For molecules with millions of lines that would increase the number
of wavelength points to be calculated (which increases the CPU time linearly)
to an unnecessarily large number.

\subsubsection{LTE tests}

In order to test the stability of the code, several tests have been
performed. All these tests take advantage of the fact that LTE has to
be restored in limiting cases. LTE is restored when the
atmosphere is dominated by collisions or if the radiation field
is described by the Planck function. LTE becomes exact when there are only
collisional rates and the radiative rates are all zero 
($C_{ij}\neq 0$, $R_{ij}\equiv 0$)
or when the source function is exactly the Planck function ($S_{\nu}=B_{\nu}$).
Both conditions can easily be applied artificially in the computer
code.
For the second test condition to be very clean and accurate, the collisional
rates are set to zero (i.e. $S_{\nu}=B_{\nu}$, $C_{ij}\equiv 0$).

All tests have been performed with a model for \teff=2700, \logg=5.0
and solar metallicity. The choice of superlevels was ``Model B''
(i.e. rotational ground states as energy boundaries).
The choice of the superlevel model does not affect this test because
a change of the superlevel model is only a change of input data,
not a change of the computer code.

The first test case was a LTE model. The departure coefficients $b_i$
remained 1.0 to an accuracy of about $10^{-7}$ for several iterations for
both test conditions.

The second test case was a LTE model, but the departure coefficients
were artificially changed under the restriction of particle conservation.
That means some fraction (10--50
state were put into another excited state.
This yields $b_i$'s of 0.5 or 0.9 for the de-populated state and
$b_i$'s of up to ten or more for the over-populated state.
The LTE situation was restored within only one iteration with an
accuracy of about $10^{-7}$ for both test conditions as well.

\subsubsection{Numerical stability}

The collisional rates couple all levels with each other.
The energy differences (and therefore, the rates) between two levels can vary
by several orders of magnitudes when comparing arbitrary
pairs of levels.
That has important numerical complications for Model~A.
In Model~A, the individual transitions that  make up one particular supertransition
do have largely varying energy differences and rates.
Since a collisional rate between two superlevels is the sum of the many individual
rates we had to assure that the numerically correct sum got evaluated.
Otherwise, the collisional rates did not obey the Boltzmann
relation \eqref{sucollratio} anymore and the LTE tests described above would fail.

This is a disadvantage of Model~A. All other models do not have this problem since
the energy differences of the transitions that make up the supertransition
are of the same order of magnitude.
In particular, Model~C can be regarded as Model~A which has been modified
in order to decrease the energy differences which then results in
numerical stability.
For future molecules we will have to assure numerical stable
summation if possible or otherwise use a different superlevel model.

\subsection{Results for a test molecule}

In the following we present results of a test molecule strongly affected
by non--LTE effects.
We used the CO molecule without accounting for collisions
with H, H$_2$ and He.
This is a very unrealistic model. However, it will be in non--LTE
in cool atmospheres. Therefore, we used it to test our
method in the non--LTE case.

The models in this section are fully converged and have
\teff=2700~K, \logg=5.0 and solar metallicity.

\subsubsection{Behavior of the departure coefficients}

\begin{figure}
\plotone{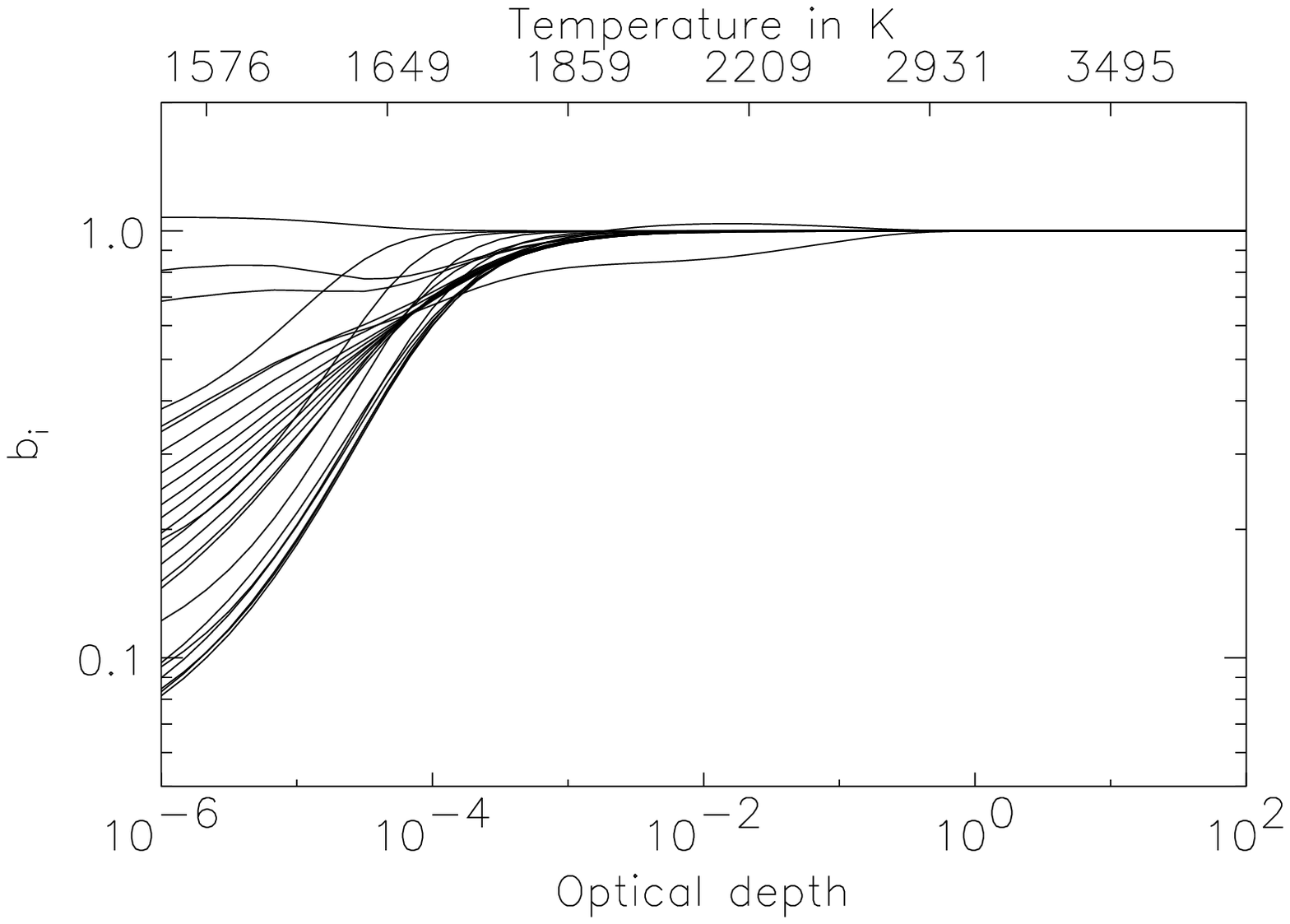}
\caption{\label{bi_quant_tcor}
The departure coefficients for a converged model with 
temperature correction. Selection method is ``Model A'', i.e. 
the vibrational quantum number. The temperature
scale on the top is the electron temperature at the respective
depth.}
\end{figure}

\begin{figure}
\plotone{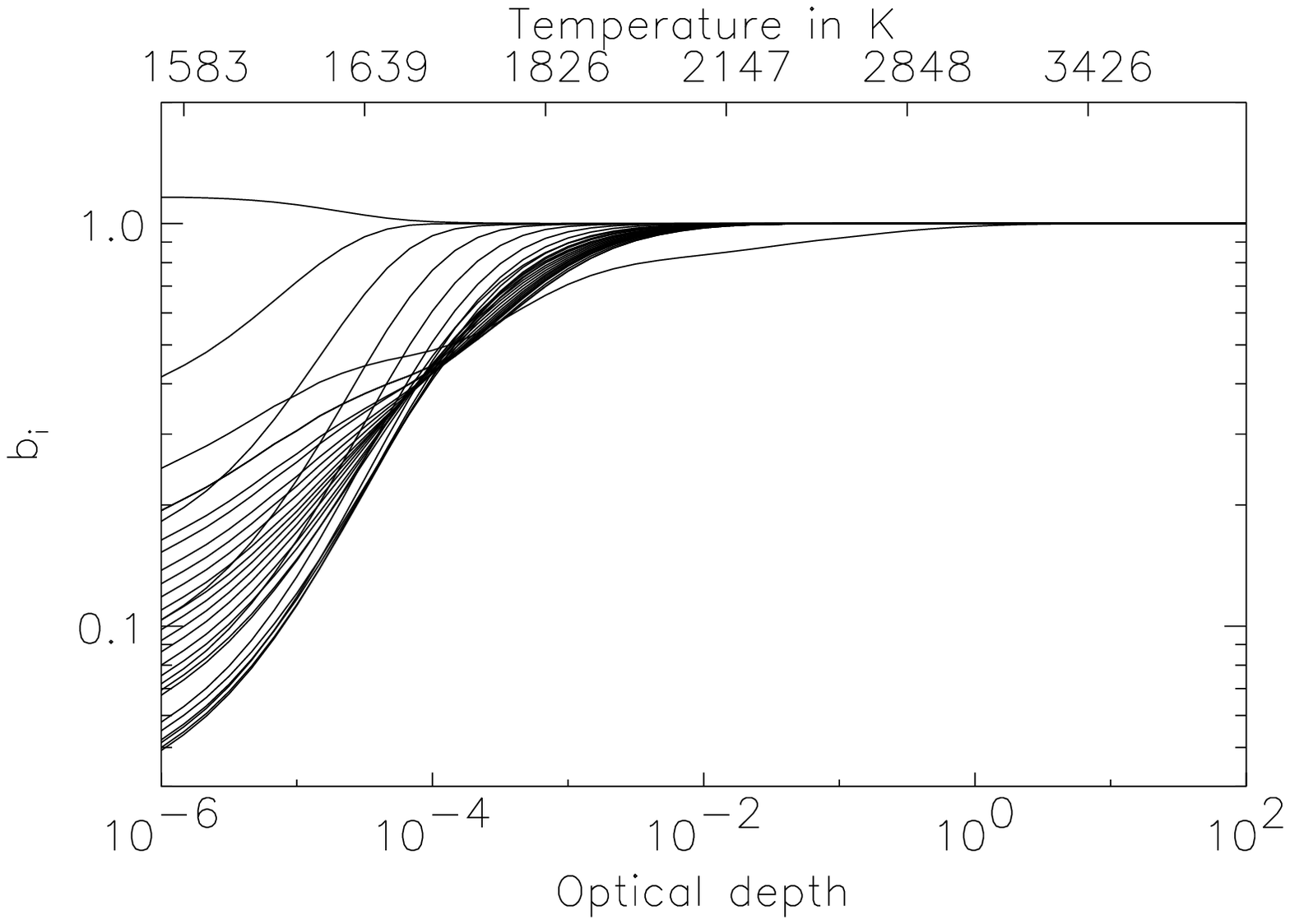}
\caption{\label{bi_head_tcor}
The departure coefficients for a converged model with 
temperature correction. Selection method is ``Model B'', i.e. 
the energy of the rotational ground states. The temperature
scale on the top is the electron temperature at the respective
depth.}
\end{figure}

\begin{figure}
\plotone{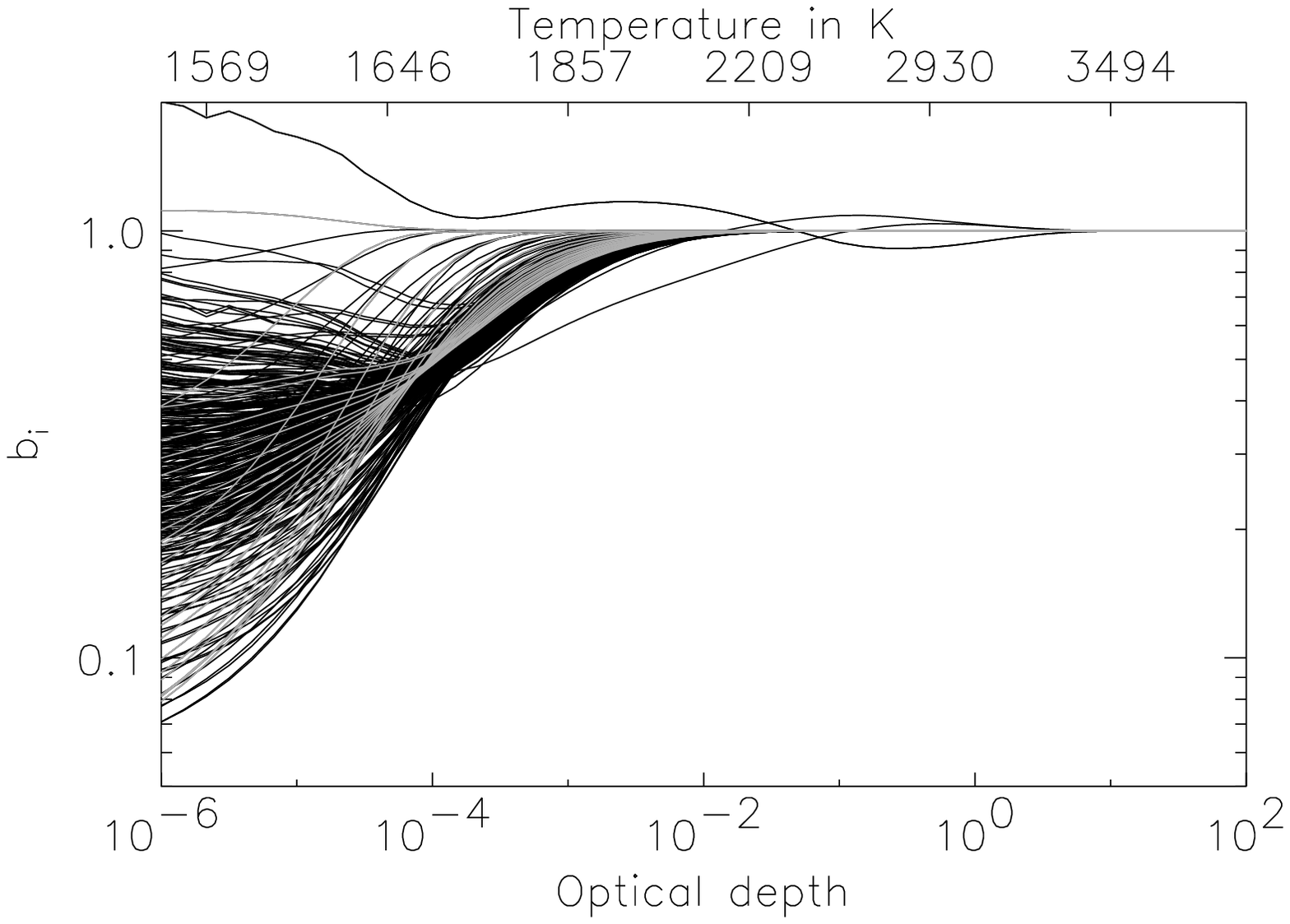}
\caption{\label{bi_square_tcor}
The departure coefficients for a converged model with 
temperature correction. Selection method is ``Model C'', i.e. 
the energy and the vibrational quantum number. The temperature
scale on the top is the electron temperature at the respective
depth.
The bright lines correspond to the superlevels
that have the lowest energies for a given vibrational
quantum number.}
\end{figure}

\begin{figure}
\plotone{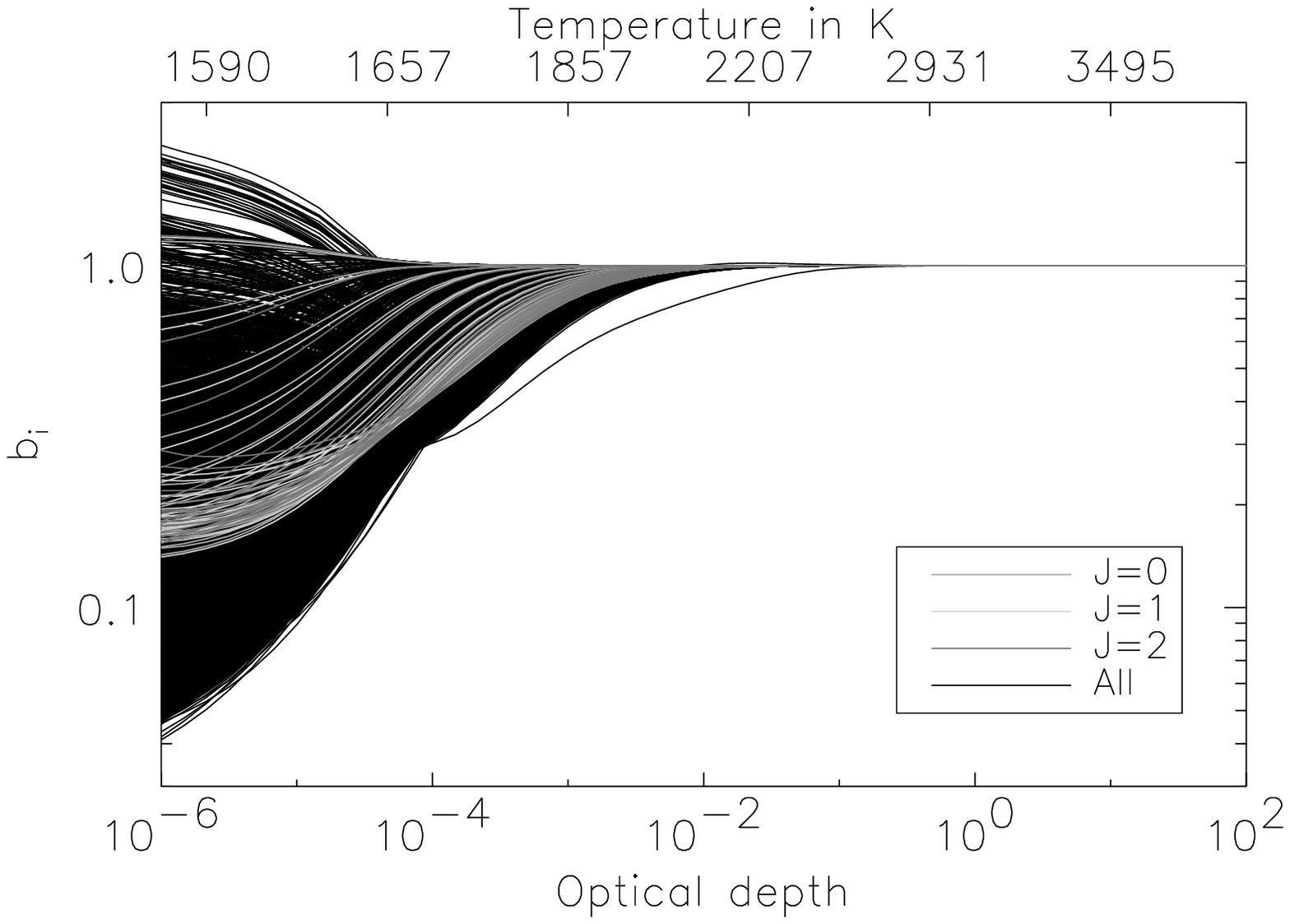}
\caption{\label{bi_real_tcor}
The departure coefficients for a converged model with 
temperature correction. The model is ``Model Z'', i.e. 
direct non--LTE. The temperature
scale on the top is the electron temperature at the respective
depth.
The levels with the lowest rotational quantum numbers are indicated in the figure.
}
\end{figure}

The departure coefficients for a converged model
can be found in Fig. \eqref{bi_quant_tcor}
for Model~A, in Fig. \eqref{bi_head_tcor} for Model~B,
in Fig. \eqref{bi_square_tcor} for Model~C and
in Fig. \eqref{bi_real_tcor} for Model~Z.
As can be seen, the $b_i$ structures are very similar
for all models.
The onset of non--LTE effects start in about the same depth of
about $\tau = 10^{-2}$ 
for all superlevel models.

Generally, it is not expected that the $b_i$ distribution
is the same for all superlevel models since the superlevel models
differ. However, Models~A and B are expected to be dominated
by the same levels. The similarity of Figs. \eqref{bi_quant_tcor} and \eqref{bi_head_tcor}
can very well be explained when we assume that Models~A and B
are dominated by the same actual levels.
This is also supported by the results from Model~C which can 
be seen in Fig.~\eqref{bi_square_tcor}.
The bright lines in Fig.~\eqref{bi_square_tcor} correspond to
the superlevels that have the lowest energies for a given vibrational
quantum number, i.e. contain the dominating levels for Models~A and B.
As one can see, the bright $b_i$ structure in Fig.~\eqref{bi_square_tcor}
is very similar to Figs. \eqref{bi_quant_tcor} and \eqref{bi_head_tcor}.
The strongest argument, however, comes from our direct non--LTE calculation.
As can be seen in Fig. \eqref{bi_real_tcor}, the levels with the lowest
rotational quantum numbers have a very similar $b_i$ structure as shown
in Figs. \eqref{bi_quant_tcor} and \eqref{bi_head_tcor}.
That means that, indeed, the superlevels are dominated by the levels
with the lowest rotational quantum numbers (or equivalently the
lowest energies) for a given vibrational quantum number
and all our superlevel models produce consistent $b_i$ structures.

\subsubsection{Effects on the spectrum}

Detailed high resolution spectra have been calculated in the
spectral regions where CO is most prominent,
i.e. between 2 and 2.4~\micron\ and between
4.3 and 6~\micron.
The $\Delta\nu = 2$ band between 2 and 2.4~\micron\
did not show any significant changes, whereas the $\Delta\nu = 1$
band between 4.3 and 6~\micron\ did.

\begin{figure}
\plotone{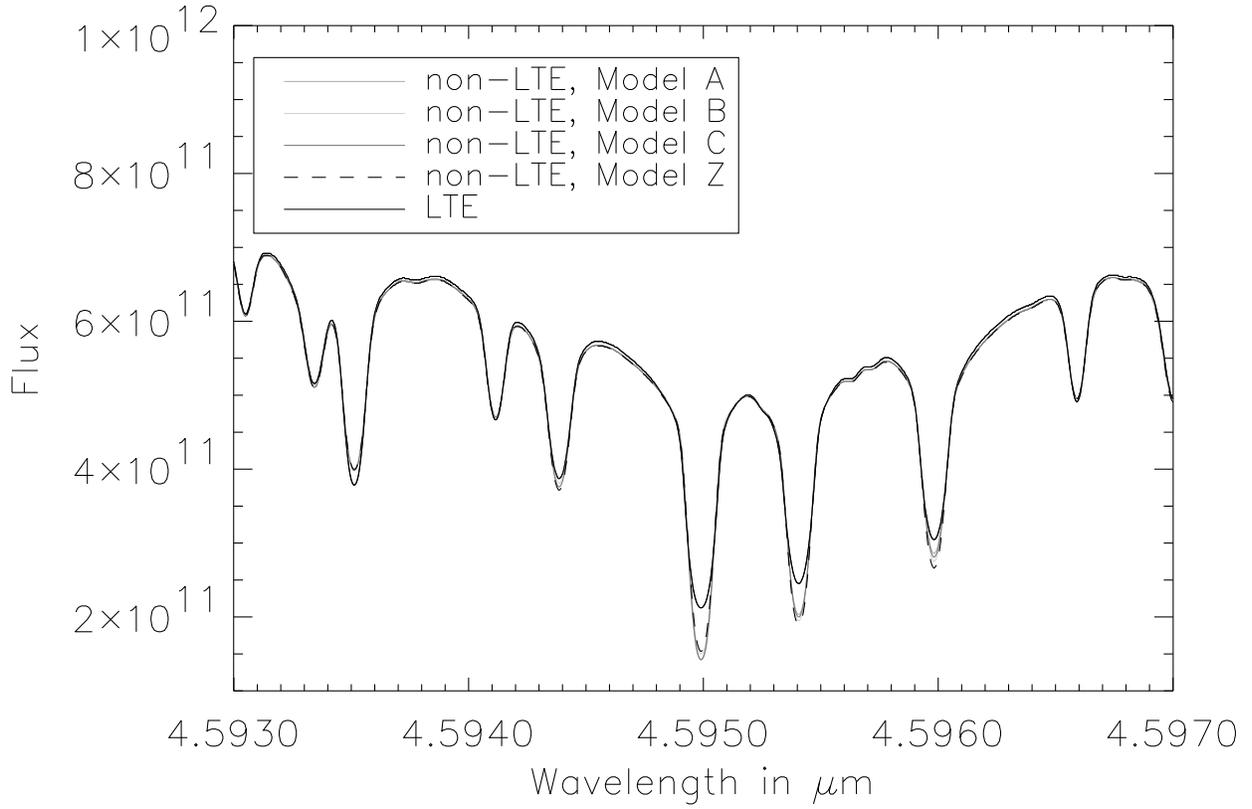}
\caption{\label{hires_head}
High resolution model spectra for an
arbitrary region in the $\Delta \nu=1$ band between 4.3 and 6~\micron.
The LTE spectrum and the non--LTE spectra with the different
models for the superlevel are indicated in the figure.
All non--LTE spectra lie on top of each other.
This spectrum is calculated with a very high resolution
to demonstrate the effects on the individual lines.
}
\end{figure}

A high resolution spectrum of a region in the $\Delta\nu = 1$ 
is shown in Fig.~\eqref{hires_head} for an LTE model and non--LTE models
with the various superlevel models discussed here.
As can be seen, all the non--LTE models, including the direct non--LTE
calculation, produce deeper lines
and are all identical within the accuracy of the calculations and
the plots.
Fig.~\eqref{hires_head} is only an example of that band. We note that
all CO lines are deeper for our non--LTE calculations and,
therefore, the whole $\Delta\nu = 1$ band is deeper.

It is important that
the spectra are the same for every superlevel model that is chosen
under the correct physical assumptions.
If we want to use the superlevel approximation we have
to make sure that the physical results are independent
of the approximation.
But most importantly,
the direct non--LTE and our superlevel models produce
the same result
for the spectrum.

The presence of non--LTE effects in the $\Delta\nu = 1$ band
also means that this band forms in regions where
deviations from LTE are present, i.e. outward of $\tau \approx 10^{-2}$.
The $\Delta\nu = 2$ band on the other hand has to form in regions deeper
than $\tau \approx 10^{-2}$.
This explains that no non--LTE effects
are visible in the spectral region between 2 and 2.4~\micron.

\subsection{Results for CO}

All models presented here have \teff=2700~K, \logg=5.0 and solar
metallicity. This roughly describes a typical M~8 dwarf.
The very low effective temperature has been deliberately chosen
since non--LTE effects are expected to be largest for low
temperatures.
The models have been calculated using the superlevel models
described above and are fully converged with regard to the
departure coefficients
and to the temperature structure.

\subsubsection{Behavior of the departure coefficients}

\begin{figure}
\plotone{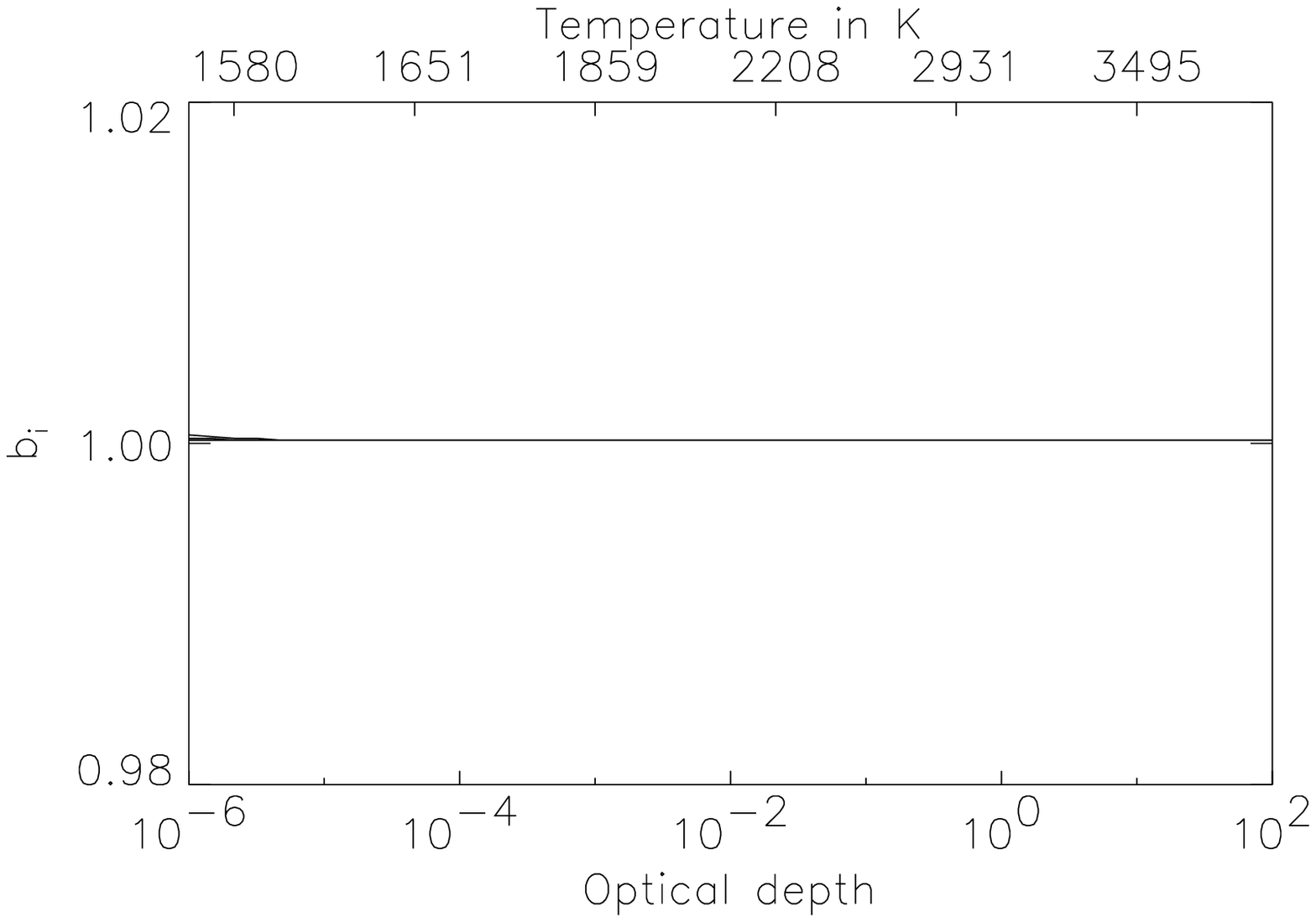}
\caption{\label{newcoll_bi_quant_tcor}
The departure coefficients for a converged model with 
temperature correction. Selection method is ``Model A'', i.e. 
the vibrational quantum number. The temperature
scale on the top is the electron temperature at the respective
depth.}
\end{figure}

\begin{figure}
\plotone{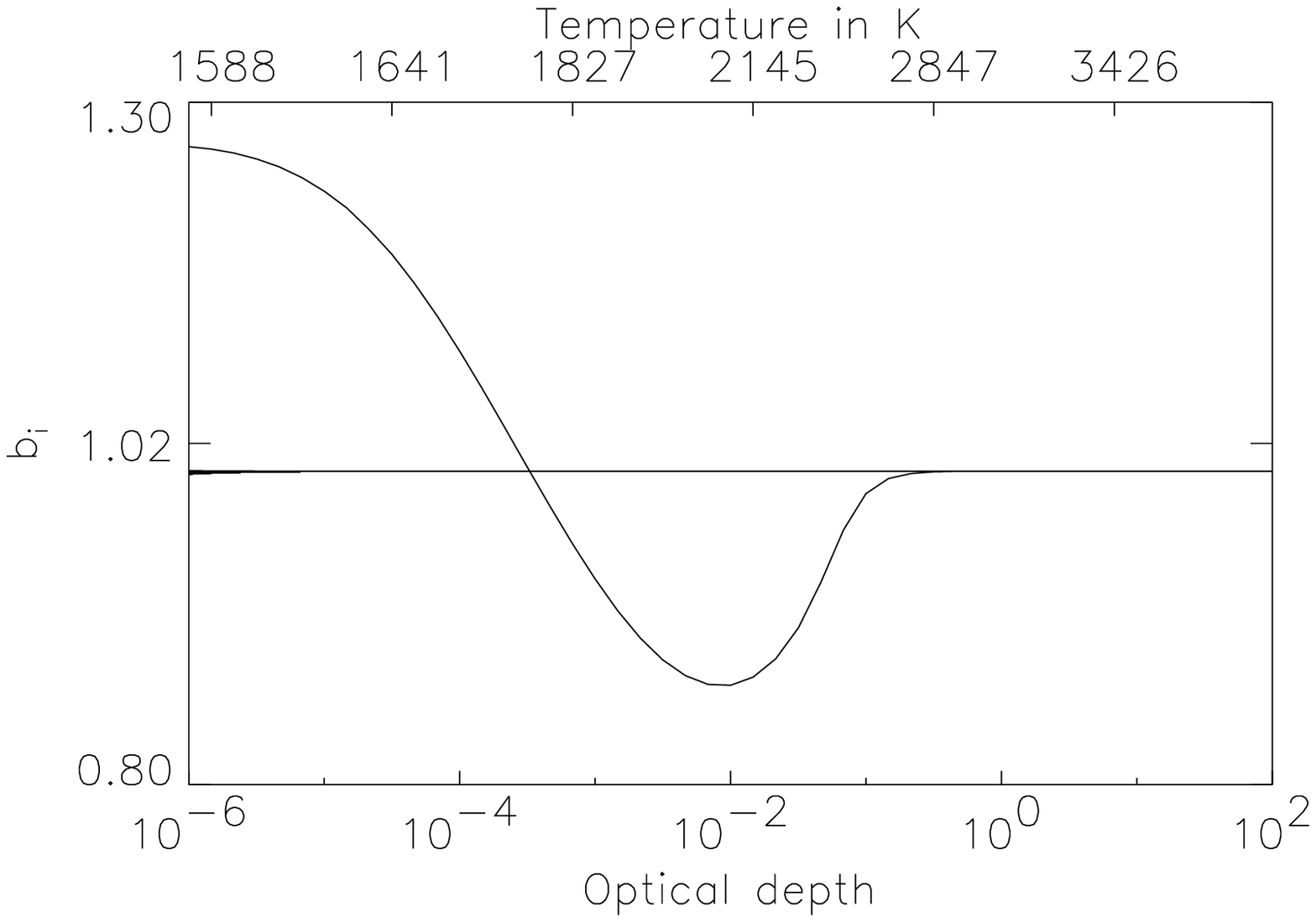}
\caption{\label{newcoll_bi_head_tcor}
The departure coefficients for a converged model with 
temperature correction. Selection method is ``Model B'', i.e. 
the energy of the rotational ground states. The temperature
scale on the top is the electron temperature at the respective
depth.}
\end{figure}

\begin{figure}
\plotone{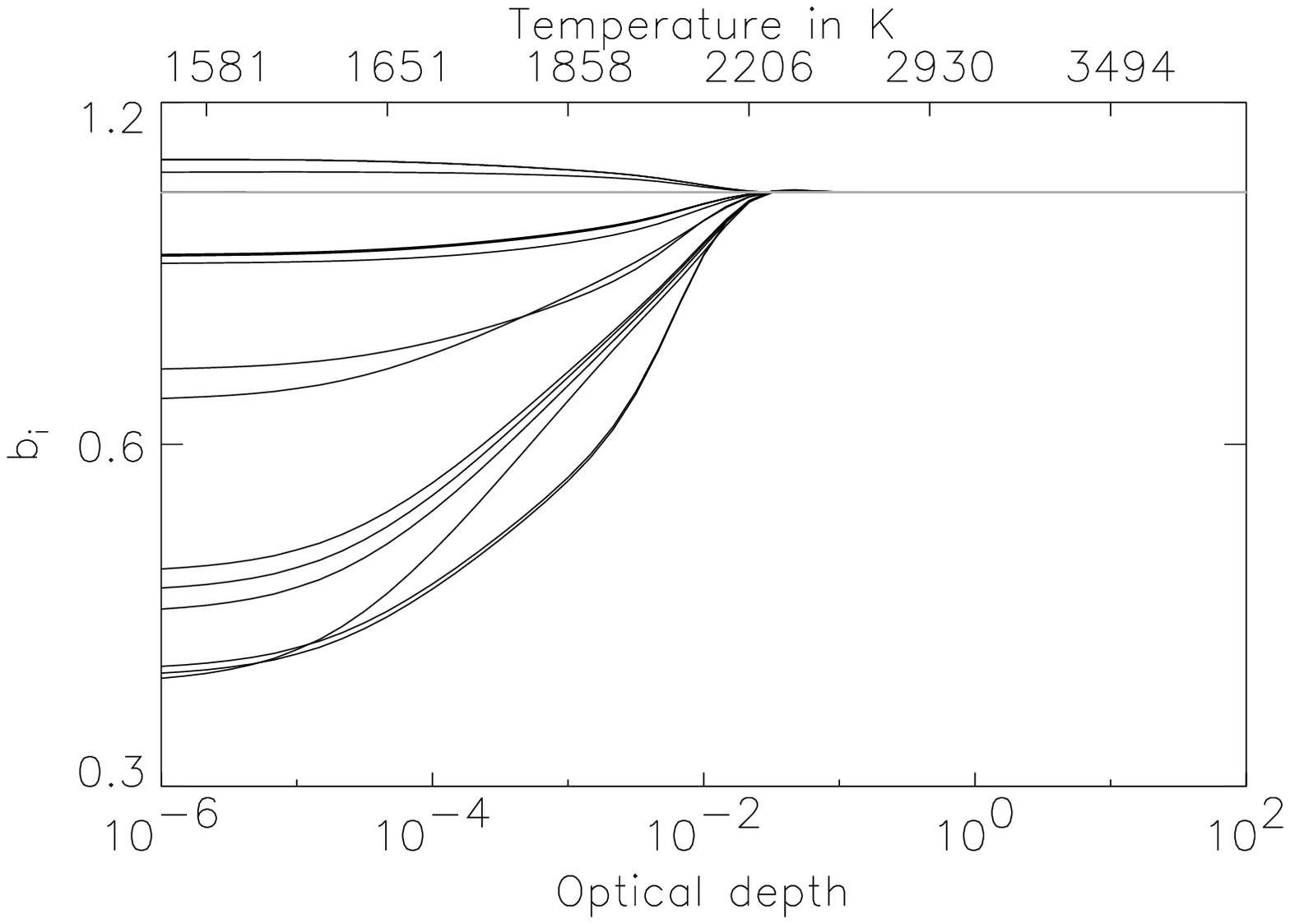}
\caption{\label{newcoll_bi_square_tcor}
The departure coefficients for a converged model with 
temperature correction. Selection method is ``Model C'', i.e. 
the energy and the vibrational quantum number. The temperature
scale on the top is the electron temperature at the respective
depth.
The bright lines correspond to the superlevels
that have the lowest energies for a given vibrational
quantum number.}
\end{figure}

\begin{figure}
\plotone{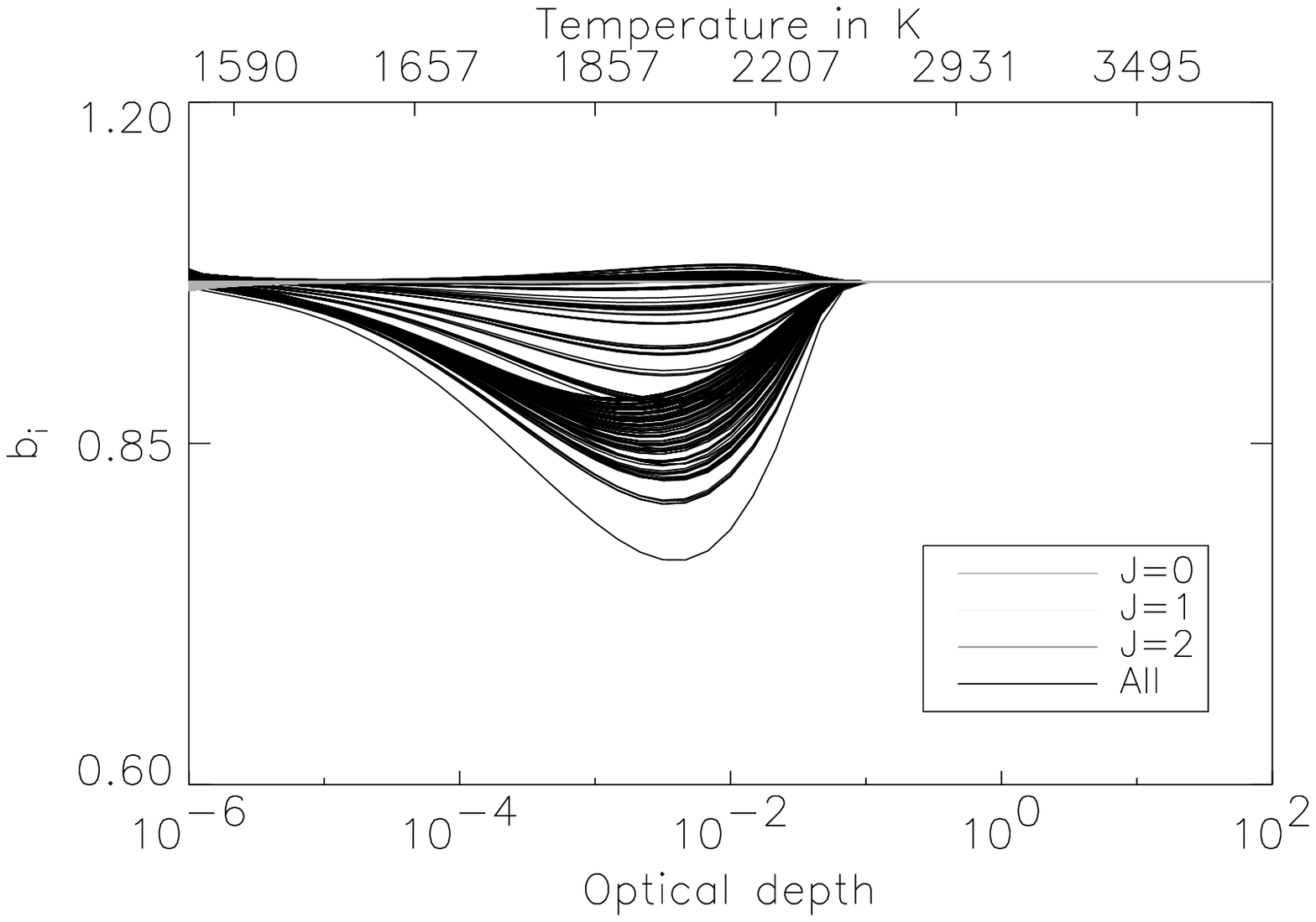}
\caption{\label{newcoll_bi_real_tcor}
The departure coefficients for a converged model with 
temperature correction. The model is ``Model Z'', i.e. 
direct non--LTE. The temperature
scale on the top is the electron temperature at the respective
depth.
The levels with the lowest rotational quantum numbers are indicated in the figure.
}
\end{figure}

The departure coefficients for a realistic CO calculation
are shown in Figs. 
\eqref{newcoll_bi_quant_tcor}
\eqref{newcoll_bi_head_tcor}
\eqref{newcoll_bi_square_tcor}
and
\eqref{newcoll_bi_real_tcor}
for Models~A, B, C and Z, respectively.
The plots show that CO stays in LTE when accounting for
collisions with H, He and H$_2$.
The levels that deviate significantly from LTE correspond
to the levels with very high excitation energies and
have negligible effects.

It is very important to note, that we can confirm the previous 
calculations that found LTE for CO not only for
approximative calculations but even if we do a
direct non--LTE calculation.

\subsubsection{The rate coefficients}

\begin{figure}
\epsscale{0.85}
\plotone{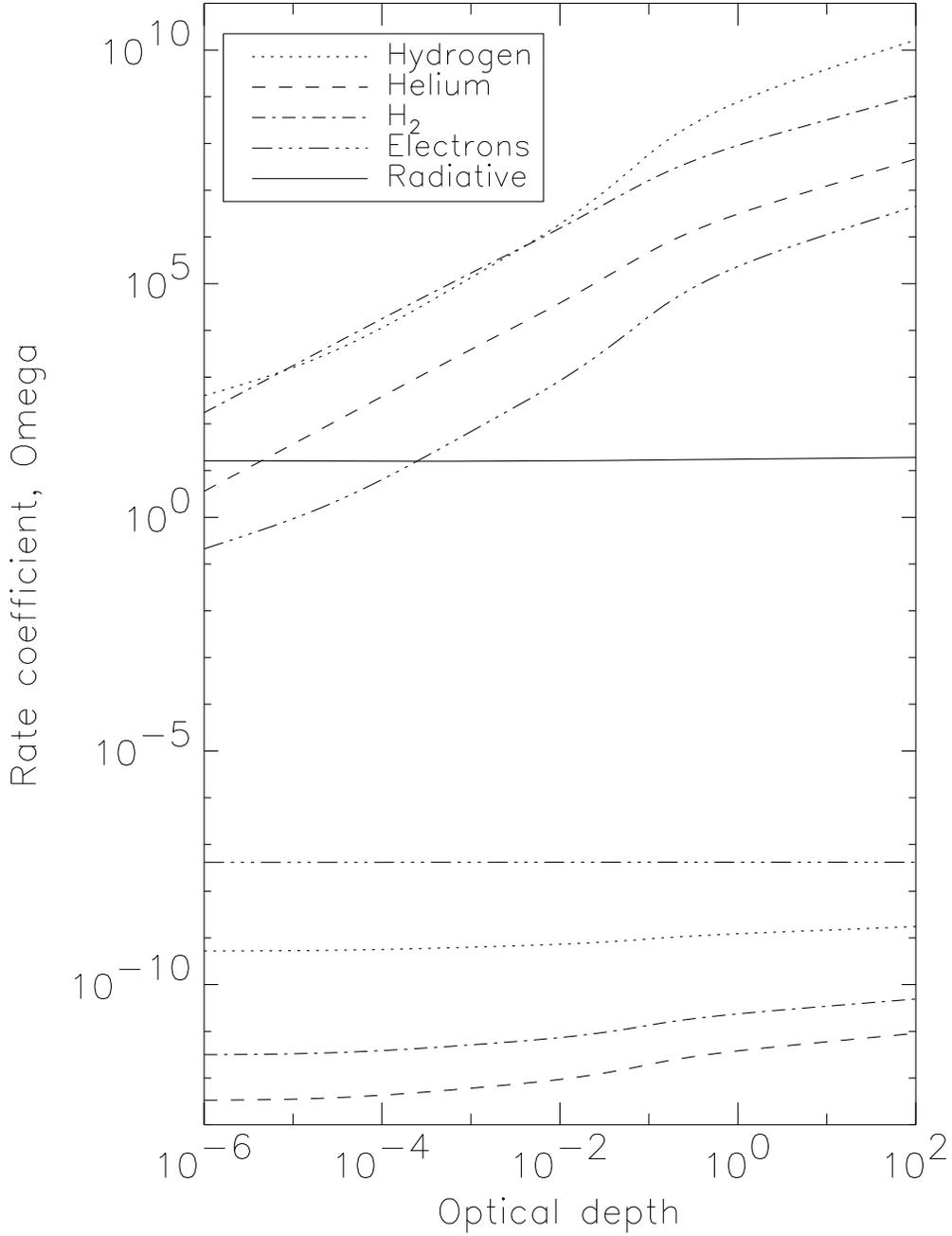}
\caption{\label{rates}
The collisional de-excitation rate coefficients, the radiative emission coefficients and
the $\Omega$ factors 
for the transition between superlevels number 7 and 5 in Model~B
(superlevel number 0 is the superlevel with the lowest excitation
energies).
The collisional partners are indicated in the figure.
The top part of the figure are collisional rate coefficients, the bottom part
are the $\Omega$ factors.
The units are \phoenix\ internal arbitrary units,
but comparable among each other.
The $\Omega$ factors have to be multiplied with the densities of the collisional
partners to obtain the rate coefficients.
}
\end{figure}

We also compared the collisional rate coefficients due to collisions with electrons,  H, He and H$_2$
with the radiative rate coefficients.
As a typical example we show the de-excitation rate coefficients $R_{75}$ for Model~B in figure \eqref{rates}.
These are the de-excitations between the superlevels defined by the vibrational
quantum numbers 7 and 5 and the respective energy bands.
As can be seen the rate coefficients due to electronic collisions are the smallest collisional
rate coefficients.
The radiative rate coefficients are generally smaller than the collisional rate coefficients,
which is pushing the system in LTE.

We also compared the density independent parts of the rate coefficients $\Omega$ for the
different collision partners. As can be seen, the electronic $\Omega$'s are the strongest,
followed by H. However, the dominating factor for the resulting rate coefficients
are the respective densities.
The two largest rate coefficients are the one for H because of its large cross section
and H$_2$ because of its large number density.

The cross sections for collisions of  H, He and H$_2$ with CO are known to be
uncertain \citep{ayres89}.
However, our results indicate that the rate coefficients for H, He and H$_2$ would have to be
several orders of magnitudes smaller to see non--LTE effects for CO.
This is in good agreement with all the previous studies.

\section{Conclusions}
\label{concs}

We have developed a method to treat molecules in non--LTE.
This method uses the superlevel formalism which reduces the
number of levels for a model molecule by a factor of hundred
or more and thereby reducing the rank of the system of rate
equations.
The important difference with respect to other implementations of
superlevels --- like in the work of \citet{anderson89},
\citet{dreizler93} or \citet{hubeny95} --- is the
treatment of line opacities. The original accuracy of
opacity treatment implemented in the atmosphere code
\phx\ is still preserved by taking advantage
of the direct opacity sampling (dOS).
This treats the line opacity as exactly as possible
and allows us to also use this also for the radiative
line rates.

We used CO as an example and showed how it is possible
to create superlevel models that 
approximate the direct calculations very well.
The most important result we can draw from CO
is the fact that we successfully divided
the CO molecule into a superlevel model molecule.
The results we obtained are independent of the
choice of the superlevel model and are identical
to the results a direct non--LTE calculation
of CO which is the largest non--LTE calculation
of a single species ever.
Therefore,
we demonstrated that our superlevel method
is a powerful and accurate method to calculate molecular
non--LTE problems.
For physical reasons Model~A has to be recommended
as long as the numerical stability is guaranteed.

For CO itself we can confirm the results from
the previous studies by, e.g., \citet[][and references therein]{ayres89}
that CO is in LTE in cool stellar atmospheres.

In future work we will apply our method to the most important, but much
more complicated molecules TiO and H$_2$O.
The results from CO suggest that it does not
matter much which superlevel we choose as long
as it is physically reasonable and numerically stable.
Therefore, it might be tempting to use Model~A 
for TiO and H$_2$O since Model~A is the most physical
and gives correct results.
However, we will investigate several
superlevel models also for TiO and H$_2$O which
will be similar to Model~B and C or variations
thereof. Thus we can assure the physical
correctness, the numerical stability and the necessary decrease of the
system of rate equations.

For molecules other than CO it will be important
to also include good cross sections for continuum processes. Continuum
processes for CO could be ignored, since CO is the most
stable molecule. However, when calculating molecules
like TiO or H$_2$O it will be important to
allow for ionization and recombination, and,
most importantly, for dissociation and molecule generation.
In many cases
dissociation will dominate over ionization.
This will couple the atomic equation of state and
the molecular equation of state by sources and
sinks of atoms and ions and production chains
for molecules have to be calculated simultaneously.
Therefore, the logical consequence is to develop 
a formalism for non local chemical equilibrium (NLCE).
NLCE will be important for molecules with low
dissociation energies like TiO. Non--LTE effects
on \ion{Ti}{1} lines have already been presented
in \citet{tinlte}, therefore, it will be necessary
to treat TiO, \ion{Ti}{1} and possibly \ion{O}{1} simultaneously 
in non--LTE.

Whereas
pure ionization can
be done  within the framework of the superlevel formalism,
the inclusion of NLCE will require further technical
developments. NLCE will not affect the rate equations
by increasing the number of levels but by
coupling to other species.
When developing a NLCE formalism, it will be based on the superlevel
formalism.
the superlevel formalism for molecules as presented
in this work for very stable molecules like CO.

Finally, since it is now possible to calculate CO with a lot 
of levels in non--LTE, these methods can now be applied to fields
other than M~dwarf atmospheres. 
One application will be to address the problem of the
strong CO bands in the solar spectrum as already mentioned.

\acknowledgements

This work was supported in part by NASA ATP grant NAG 5-8425 and LTSA grant NAG
5-3619, as well as NASA/JPL grant 961582 and by NSF grant AST-9720704 to the
University of Georgia.  This work was supported  in part by the P\^ole
Scientifique de Mod\'elisation Num\'erique at ENS-Lyon.  Some of the
calculations presented in this paper were performed on the IBM SP and the SGI
Origin 2000 of the UGA UCNS and on the IBM SP of the San Diego Supercomputer
Center (SDSC), with support from the National Science Foundation, and on the
Cray T3E of the NERSC with support from the DoE.
A.S. acknowledges furthermore support from DFG grant 1053/8-1.
E.B. acknowledges support from NSF grant AST 973140 and NASA grant NAG 5-3505.

We want to thank the referee, T. Ayres, for his very helpful comments.
A.S. wants to thank J. Krautter, S. Starrfield and I. Appenzeller who
made this work possible and supported it.
We also thank F. Allard for her work on \phx.

\bibliography{}

\begin{thebibliography}{38}
\expandafter\ifx\csname natexlab\endcsname\relax\def\natexlab#1{#1}\fi

\bibitem[{Allard \& Hauschildt(1995)}]{MDpap}
Allard, F. \& Hauschildt, P.~H. 1995, ApJ, 445, 433

\bibitem[{{Allard} {et~al.}(1997){Allard}, {Hauschildt}, {Alexander}, \&
  {Starrfield}}]{ara97}
{Allard}, F., {Hauschildt}, P.~H., {Alexander}, D.~R., \& {Starrfield}, S.
  1997, \araa, 35, 137

\bibitem[{Allard {et~al.}(1994)Allard, Hauschildt, Miller, \&
  Tennyson}]{h2olet}
Allard, F., Hauschildt, P.~H., Miller, S., \& Tennyson, J. 1994, ApJL, 426, 39

\bibitem[{Allen(1973)}]{allen_aq}
Allen, C.~W. 1973, Astrophysical Quantities, 3rd edn. (London: Athlone Press)

\bibitem[{Anderson(1989)}]{anderson89}
Anderson, L.~S. 1989, \apj, 339, 558

\bibitem[{{Ayres} \& {Wiedemann}(1989)}]{ayres89}
{Ayres}, T.~R. \& {Wiedemann}, G.~R. 1989, \apj, 338, 1033

\bibitem[{{Baron} \& {Hauschildt}(1998)}]{phxpar98}
{Baron}, E. \& {Hauschildt}, P.~H. 1998, \apj, 495, 370

\bibitem[{Drawin(1961)}]{drawin61}
Drawin, H.~W. 1961, Zs. f. Phys., 164, 513

\bibitem[{Dreizler \& Werner(1993)}]{dreizler93}
Dreizler, S. \& Werner, K. 1993, A\&A, 278, 199

\bibitem[{Glass \& Kironde(1982)}]{glass82}
Glass, G.~P. \& Kironde, S. 1982, J. Phys. Chem., 86, 908

\bibitem[{Goorvitch(1994)}]{goorCO}
Goorvitch, D. 1994, ApJS, 95, 535

\bibitem[{Goorvitch \& {Chackerian, Jr.}(1994{\natexlab{a}})}]{goorCOa}
Goorvitch, D. \& {Chackerian, Jr.}, C. 1994{\natexlab{a}}, ApJS, 91, 483

\bibitem[{Goorvitch \& {Chackerian, Jr.}(1994{\natexlab{b}})}]{goorCOb}
---. 1994{\natexlab{b}}, ApJS, 92, 311

\bibitem[{Hauschildt(1992)}]{s3pap}
Hauschildt, P.~H. 1992, JQSRT, 47, 433

\bibitem[{Hauschildt(1993)}]{casspap}
---. 1993, JQSRT, 50, 301

\bibitem[{Hauschildt {et~al.}(1997{\natexlab{a}})Hauschildt, Allard, Alexander,
  \& Baron}]{tinlte}
Hauschildt, P.~H., Allard, F., Alexander, D.~R., \& Baron, E.
  1997{\natexlab{a}}, ApJ, 488, 428

\bibitem[{Hauschildt {et~al.}(1996{\natexlab{a}})Hauschildt, Allard, Alexander,
  Schweitzer, \& Baron}]{IAU176}
Hauschildt, P.~H., Allard, F., Alexander, D.~R., Schweitzer, A., \& Baron, E.
  1996{\natexlab{a}}, in Stellar Surface Structure, ed. K.~G. Strassmeier \&
  J.~L. Linsky, I.A.U. Symposium 176 (Dordrecht: Kluwer), 539

\bibitem[{Hauschildt {et~al.}(1999)Hauschildt, Allard, \& Baron}]{NGhot}
Hauschildt, P.~H., Allard, F., \& Baron, E. 1999, ApJ, 512, 377

\bibitem[{Hauschildt {et~al.}(1997{\natexlab{b}})Hauschildt, Baron, \&
  Allard}]{phxparallel}
Hauschildt, P.~H., Baron, E., \& Allard, F. 1997{\natexlab{b}}, ApJ, 483, 390

\bibitem[{Hauschildt {et~al.}(1996{\natexlab{b}})Hauschildt, Baron, Starrfield,
  \& Allard}]{fe2effects}
Hauschildt, P.~H., Baron, E., Starrfield, S., \& Allard, F. 1996{\natexlab{b}},
  ApJ, 462, 386

\bibitem[{{Hauschildt} {et~al.}(1997){Hauschildt}, {Shore}, {Schwarz}, {Baron},
  {Starrfield}, \& {Allard}}]{novaphys97}
{Hauschildt}, P.~H., {Shore}, S.~N., {Schwarz}, G.~J., {Baron}, E.,
  {Starrfield}, S., \& {Allard}, F. 1997, \apj, 490, 803

\bibitem[{Hauschildt {et~al.}(1994)Hauschildt, Starrfield, Austin, Wagner,
  Shore, \& Sonneborn}]{cygpap}
Hauschildt, P.~H., Starrfield, S., Austin, S.~J., Wagner, R.~M., Shore, S.~N.,
  \& Sonneborn, G. 1994, ApJ, 422, 831

\bibitem[{Hubeny \& Lanz(1995)}]{hubeny95}
Hubeny, I. \& Lanz, T. 1995, ApJ, 439, 875

\bibitem[{Kutepov {et~al.}(1991)Kutepov, Kunze, Hummer, \& Rybicki}]{kutepov91}
Kutepov, A.~A., Kunze, D., Hummer, D., \& Rybicki, G. 1991, JQSRT, 46, 347

\bibitem[{Kutepov {et~al.}(1997)Kutepov, Oelhaf, \& Fischer}]{kutepov97}
Kutepov, A.~A., Oelhaf, H., \& Fischer, H. 1997, JQSRT, 57, 317

\bibitem[{Menzel \& Cilli{\'e}(1937)}]{menzel37}
Menzel, D.~H. \& Cilli{\'e}, G. 1937, \apj, 85, 88

\bibitem[{Mihalas(1978)}]{mihalas78}
Mihalas, D. 1978, Stellar Atmospheres, 2nd edn. (San Fransisco: W. H. Freeman
  and Company)

\bibitem[{Milikan(1964)}]{milikan64}
Milikan, R.~C. 1964, J. Chem. Phys., 40, 2594

\bibitem[{Partridge \& Schwenke(1997)}]{ames-water-new}
Partridge, H. \& Schwenke, D.~W. 1997, J. Chem. Phys., 106, 4618

\bibitem[{{Rybicki} \& {Hummer}(1991)}]{rh91}
{Rybicki}, G.~B. \& {Hummer}, D.~G. 1991, \aap, 245, 171

\bibitem[{Schweitzer(1995)}]{diplom}
Schweitzer, A. 1995, Master's thesis, University of Heidelberg, in german

\bibitem[{Schweitzer(1999)}]{phd}
---. 1999, PhD thesis, University of Heidelberg

\bibitem[{Schweitzer {et~al.}(1996)Schweitzer, Hauschildt, Allard, \&
  Basri}]{VB10}
Schweitzer, A., Hauschildt, P.~H., Allard, F., \& Basri, G. 1996, MNRAS, 283,
  821

\bibitem[{Schwenke(1998)}]{ames-tio}
Schwenke, D.~W. 1998, ???

\bibitem[{Thompson(1973)}]{thompson73}
Thompson, R.~I. 1973, ApJ, 181, 1039

\bibitem[{{Van Regemorter}(1962)}]{vr62}
{Van Regemorter}, H. 1962, \apj, 136, 906

\bibitem[{{Wiedemann} {et~al.}(1994){Wiedemann}, {Ayres}, {Jennings}, \&
  {Saar}}]{wiedemann94}
{Wiedemann}, G., {Ayres}, T.~R., {Jennings}, D.~E., \& {Saar}, S.~H. 1994,
  \apj, 423, 806

\bibitem[{Wijbenga \& Zwaan(1972)}]{zwaan72}
Wijbenga, J.~W. \& Zwaan, C. 1972, Solar Phys., 23, 265

\end{thebibliography}

\newpage

\end{document}